\documentclass[fleqn,usenatbib,usedcolumns]{mnras}

\usepackage{amsmath}
\usepackage{txfonts}

\usepackage[british]{babel} 
\usepackage[T1]{fontenc}
\usepackage{ae,aecompl}

\usepackage{graphicx}	
\usepackage{xspace}

\graphicspath{{./Paper_plots/}}
\newcommand{\galform}{{\sc{galform}}\xspace}
\newcommand{\grasil}{{\sc{grasil}}\xspace}
\newcommand{\cosmos}{{\sc{cosmos}}\xspace}
\newcommand{\uvista}{{\sc{u}}ltra{\sc{vista}}\xspace}
\newcommand\ddfrac[2]{\frac{\displaystyle #1}{\displaystyle #2}}

\newcommand{\Td}{T_{\rm d}}

\title[The FIR SEDs of galaxies]{The far infra-red SEDs of main sequence and starburst galaxies}

\author[W. I. Cowley et al.]{
William I. Cowley$^{1\mathrm{,}2}$\thanks{E-mail: cowley@astro.rug.nl (WIC)},
Matthieu B{\'e}thermin$^{3}$,
Claudia del P. Lagos$^{4,5}$,
Cedric G. Lacey$^{1}$,\newauthor
Carlton M. Baugh$^{1}$,
Shaun Cole$^{1}$
\\
$^{1}$Institute for Computational Cosmology, Department of Physics, University of Durham, South Road, Durham, DH1 3LE, UK\\
$^{2}$Kapteyn Astronomical Institute, University of Groningen, P.O. Box 800, 9700 AV Groningen, The Netherlands\\
$^{3}$European Southern Observatory, Karl-Schwarzschild-Str. 2, 85748 Garching, Germany\\
$^{4}$International Centre for Radio Astronomy Research (ICRAR), M468, University of Western Australia, 35 Stirling Hwy, Crawley, WA 6009,\\ Australia\\
$^{5}$Australian Research Council Centre of Excellence for All-sky Astrophysics (CAASTRO), 44 Rosehill Street Redfern, NSW 2016, Australia\\
}

\date{Accepted XXX. Received YYY; in original form ZZZ}

\pubyear{2016}

\begin{document}
\label{firstpage}
\pagerange{\pageref{firstpage}--\pageref{lastpage}}
\maketitle

\begin{abstract}
We compare observed far infra-red/sub-millimetre (FIR/sub-mm) galaxy spectral energy distributions (SEDs) of massive galaxies ($M_{\star}\gtrsim10^{10}$~$h^{-1}$~M$_{\sun}$) derived through a stacking analysis with predictions from a new model of galaxy formation.  The FIR SEDs of the model galaxies are calculated using a self-consistent model for the absorption and re-emission of radiation by interstellar dust based on radiative transfer calculations and global energy balance arguments.  Galaxies are selected based on their position on the specific star formation rate (sSFR) - stellar mass ($M_{\star}$) plane.  We identify a main sequence of star-forming galaxies in the model, i.e. a well defined relationship between sSFR and $M_\star$, up to redshift $z\sim6$. The scatter of this relationship evolves such that it is generally larger at higher stellar masses and higher redshifts. There is remarkable agreement between the predicted and observed average SEDs across a broad range of redshifts ($0.5\lesssim z\lesssim4$) for galaxies on the main sequence.  However, the agreement is less good for starburst galaxies at $z\gtrsim2$, selected here to have elevated {sSFRs}$>10\times$ the main sequence value.  We find that the predicted average SEDs are robust to changing the parameters of our dust model within physically plausible values.  We also show that the dust temperature evolution of main sequence galaxies in the model is driven by star formation on the main sequence being more burst-dominated at higher redshifts.    
\end{abstract}



\section{Introduction}
\label{sec:Introduction}
Interstellar dust plays an important role in observational probes of galaxy formation and evolution.  It forms from metals produced by stellar nucleosynthesis which are then ejected by stellar winds and supernovae into the interstellar medium (ISM), where a fraction ($\sim$~$30-50$~per~cent, e.g. Draine \& Li \citeyear{DL07}) condense into grains.  These grains then absorb stellar radiation and re-emit it at longer wavelengths.  Studies of the extragalactic background light have found that the energy density of the cosmic infra-red background (CIB, $~\sim10-1000$~$\muup$m) is similar to that found in the UV/optical/near infra-red \citep[e.g.][]{HauserDwek01,Dole06}, suggesting that much of the star formation over the history of the Universe has been obscured by dust.  Thus understanding the nature of dust and its processing of stellar radiation is crucial to achieve a more complete view of galaxy formation and evolution.

Observations suggest that the majority of star formation over the history of the Universe has taken place on the so-called main sequence (MS) of star-forming galaxies, a tight correlation between star formation rate (SFR) and stellar mass ($M_{\star}$) that is observed out to $z\sim4$, with a $1\sigma$ scatter of $\sim0.3$~dex (e.g. Elbaz et al. \citeyear{Elbaz07}; Karim et al. \citeyear{Karim11}; Rodighiero et al. \citeyear{Rodighiero11}; for theoretical predictions see also Mitchell et al. \citeyear{Mitchell14}).  This is thought to result from the regulation of star formation through the interplay of gas cooling and feedback processes.  Galaxies that have elevated SFRs (typically by factors $\sim4-10$) relative to this main sequence are often referred to as starburst galaxies (SB) in observational studies.  In contrast to the secular processes thought to drive star formation on the MS, the elevated SFRs in SB galaxies are thought to be triggered by some dynamical process such as a galaxy merger or disc instability.  

The SFRs in these galaxies are usually inferred from a combination of UV and IR photometry and thus a good understanding of the effects of dust in these galaxies is important.  However, understanding the dust emission properties of these galaxies is both observationally and theoretically challenging.  Observationally, an integrated FIR SED for the whole galaxy is required to give an indication of the luminosity from young stars that is absorbed and re-emitted by the dust.  As is discussed in the following paragraph, at far infra-red/sub-mm (FIR/sub-mm) wavelengths broadband photometry is complicated by issues such as confusion due to the coarse angular resolution of single dish telescopes at these wavelengths.  Evolutionary synthesis models are then required to convert the infra-red luminosity derived from the observed photometry into a star formation rate \citep[e.g.][]{Kennicutt98}.  However, these must make assumptions about the star formation history of the galaxy and the stellar initial mass function (IMF).  Various models for dust emission and galaxy SEDs, that often include evolutionary synthesis and make further assumptions about the composition and geometry of the dust, can be fitted to the observed FIR/sub-mm photometry (e.g. Silva et al. \citeyear{Silva98}; Draine \& Li \citeyear{DL07}; da Cunha, Charlot \& Elbaz \citeyear{daCunha08}) to give estimates for physical dust properties such as the dust temperature ($T_{\rm d}$) and dust mass ($M_{\rm dust}$).

As mentioned above, a significant difficulty with FIR/sub-mm imaging surveys of high-redshift galaxies is the coarse angular resolution of single-dish telescopes at these long wavelengths [$\sim20$ arcsec full width half maximum (FWHM)].  This, coupled with the high surface density of detectable objects, means that imaging is often confusion-limited and that only the brightest objects (with the highest SFRs) can be resolved as point sources above the confusion background \citep[e.g.][]{Nguyen10}.  These resolved galaxies either form the massive end of the MS or have elevated SFRs relative to the MS and are thus defined as starburst galaxies (SB).  

At $z\sim2$, MS galaxies have SFRs high enough to be resolved in \emph{Herschel}\footnote{\url{http://sci.esa.int/herschel/}} imaging only if they have large stellar masses ($M_{\star}\gtrsim10^{10.5}$~$h^{-1}$~M$_{\sun}$) whereas SB galaxies with stellar mass approximately an order of magnitude lower can still be resolved \citep[e.g.][]{Gruppioni13}.  For less massive MS galaxies and galaxies at higher redshifts, as  it is not possible to individually resolve a complete sample of galaxies, stacking techniques have been developed to overcome the source confusion and derive average FIR/sub-mm SEDs for different samples \citep[e.g.][]{Magdis12,Magnelli14,Santini14,Bethermin15}.  These studies typically begin with a stellar mass selected sample and stack available FIR/sub-mm imaging at the positions of these galaxies, in bins of stellar mass and redshift.

An early study using this technique, \cite{Magdis12},  fitted the dust model of \cite{DL07} to stacked FIR/sub-mm SEDs of $M_{\star}\gtrsim3.6\times10^{9}$~$h^{-1}$~M$_{\sun}$ galaxies at $z\sim1$ and $z\sim2$.  The Draine and Li model describes interstellar dust as a mixture of polyaromatic hydrocarbon molecules (PAHs), as well as carbonaceous and amorphous silicate grains, with the fraction of dust in PAHs determined by the parameter $q_{\rm PAH}$. The size distributions of these species are chosen such that observed extinction laws in the Milky Way, Large Magellanic Cloud and the Small Magellanic Cloud are broadly reproduced.  Dust is assumed to be heated by a radiation field with constant intensity, $U_{\rm min}$, with some fraction, $\gamma$, being exposed to a radiation field ranging in intensity from $U_{\rm min}$ to $U_{\rm max}$, representing dust enclosed in photodissociation regions.  This model thus provides a best fitting value for the total dust mass, $U_{\rm min}$, $\gamma$ and $q_{\rm PAH}$.  The resulting average radiation field $\langle U\rangle$ is strongly correlated with average dust temperature.   

Magdis et al. found that the dust temperatures of MS galaxies increases with redshift.  \cite{Bethermin15} extended this analysis to $z\sim4$ by stacking on a stellar mass-selected sample ($M_{\star}>2.1\times10^{10}$~$h^{-1}$~M$_{\sun}$) of galaxies derived from \uvista data \citep{Ilbert13} in the \cosmos field.  B\'{e}thermin et al. found, similarly to Magdis et al., that the dust temperatures of MS galaxies increases with redshift.  From fitting the \cite{DL07} dust model to their stacked SEDs, Bethermin et al. found a strong increase in the mean intensity of the radiation field, $\langle U\rangle$, which is strongly correlated with $\Td$, for MS galaxies at $z\gtrsim2$. This led these authors to suggest a break to the fundamental metallicity relation \citep[FMR,][]{Mannucci10}, which connects gas metallicity to SFR and stellar mass, and is observed to be redshift independent for $z\lesssim2$.  This break has the effect of reducing the gas metallicity (and hence dust mass) at a given stellar mass for $z\gtrsim2$.  This results in hotter dust temperatures than is implied by simply extrapolating the FMR from lower redshifts.  Bethermin et al. also performed their stacking analysis on a sample of SB galaxies, finding no evidence for dust temperature evolution with redshift for these galaxies, and that they have a similar temperature to the $z\sim2$ main sequence sample.               

Here we compare predictions from a state-of-the-art semi-analytic model of hierarchical galaxy formation within the $\Lambda$CDM paradigm \citep[\galform,][hereafter L16]{Lacey16}  to the observations presented in \cite{Bethermin15}.  B\'{e}thermin et al. also compared their inferred dust-to-stellar mass ratios and gas fractions directly with those predicted by the \galform models of L16 and Gonzalez-Perez et al. (\citeyear{vgp14}, hereafter GP14).  Here, we extend this by comparing the FIR/sub-mm SEDs directly and inferring physical properties for both the observed and simulated galaxies in a consistent manner.  In the model, the FIR/sub-mm emission is calculated by solving the equations of radiative transfer for dust absorption in an assumed geometry; and by applying energy balance arguments for dust emission to solve for the dust temperature, assuming the dust emits as a modified blackbody. Importantly, this means that the dust temperature is a prediction of the model and not a free parameter.  The L16 model can reproduce an unprecedented range of observational data, notably the optical and near infra-red luminosity functions of the galaxy population from $z=0$ to $z\sim3$, and the FIR/sub-mm number counts and redshift distributions \citep[from $250$ to $1100$~$\muup$m, L16, see also][]{Cowley15}.  An important feature of the model is that it incorporates two modes of star formation, a quiescent mode which is fuelled by gas accretion onto a galactic disc and a burst mode in which a period of enhanced star formation is triggered by a dynamical process, either a galaxy merger or disc instability.  

In order to avoid confusion with the definition of starburst arising from a galaxy's position on the sSFR-$M_{\star}$ plane relative to the main sequence, throughout this paper we will refer to populations of galaxies selected in this manner as MS, if they lie on the locus of the star-forming main sequence, or SB, if they are found at elevated SFRs relative to this locus.  Additionally,  we will refer to populations of galaxies selected according to the \galform star formation mode which is dominating their current total SFR as quiescent mode dominated and burst mode dominated populations respectively.

This paper is structured as follows: In Section~\ref{sec:Model} we describe the galaxy formation model and the model for the reprocessing of stellar radiation by dust; in Section~\ref{sec:Results} we present our results\footnote{Some of the results presented here will be made available at \url{http://icc.dur.ac.uk/data/}.  For other requests please contact the first author.}, which include a detailed comparison with the observed stacked FIR/sub-mm SEDs of \cite{Bethermin15}.  We conclude in Section~\ref{sec:conclusion}.  Throughout we assume a flat $\Lambda$CDM cosmology with cosmological parameters consistent with the $7$ year \emph{Wilkinson Microwave Anisotropy Probe} (\emph{WMAP7}) results \citep{Komatsu11} i.e. ($\Omega_{0}$, $\Lambda_{0}$, $h$, $\Omega_{\rm b}$, $\sigma_{8}$, $n_{\rm s}$) $=$ ($0.272$, $0.728$, $0.704$, $0.0455$, $0.81$, $0.967$), to match those used in L16.

\section{The Theoretical Model}
\label{sec:Model}
In this section we introduce our model, which combines a cosmological $N$-body simulation of dark matter, a state-of-the-art semi-analytic model of galaxy formation and a simple model for the reprocessing of stellar radiation by dust in which the dust temperature is calculated based on radiative transfer and global energy balance arguments.  We give an overview of the model and describe some aspects that are particularly relevant to this study in the following subsections.  For further details we refer the reader to L16. 
\subsection{GALFORM}
The Durham semi-analytic model of hierarchical galaxy formation, \galform, was introduced in \cite{Cole00}, building on ideas outlined by \cite{WhiteRees78}, \cite{WhiteFrenk91} and \cite{Cole94}.  Galaxy formation is modelled \emph{ab initio}\footnote{In the sense that the galaxy formation calculation follows all of the main physical processes from high redshift ($z\gtrsim20$).}, beginning with a specified cosmology and a linear power spectrum of density fluctuations and ending with predicted galaxy properties at different redshifts.

Galaxies are assumed to form from baryonic condensations within the potential well of a dark matter halo, with their subsequent evolution controlled in part by the merging history of the halo.  These halo merger trees can be calculated using either a Monte Carlo technique following the extended Press-Schechter formalism \citep[e.g.][]{PCH08}, or extracted directly from a dark matter only $N$-body simulation \citep[e.g.][]{Helly03,Jiang14}.  Here we use halo merger trees derived from a Millennium-type dark matter only simulation \citep{Guo13}, but with cosmological parameters consistent with the \emph{WMAP7} results \citep{Komatsu11}.

In \galform, the baryonic processes thought to be important for galaxy formation are included as a set of continuity equations which track the exchange of mass and metals between the stellar, cold disc gas and hot halo gas components in a given halo.  An example of this for the mass of metals in the cold gas component is
\begin{equation}
\dot{M}_{\rm cold}^{\rm Z}=Z_{\rm hot}\dot{M}_{\rm cool}+[p-(1-R+\beta)Z_{\rm cold}]\psi{\rm,}
\label{eq:mzcolddot}
\end{equation}
(see equation~$26$ in L16).  Here $p$, the \emph{yield} (the fraction of the initial mass of a stellar population that is synthesised into metals and then ejected), and $R$, the \emph{returned fraction}(the fraction of the initial mass of a stellar population that is returned to the ISM by mass loss from dying stars assuming instantaneous recycling) depend on the choice of the stellar IMF.  The ``mass-loading'' factor $\beta$ is the ratio of the rate at which cold gas mass is ejected from the galaxy and the halo by supernova feedback ($\dot{M}_{\rm eject}$) to the SFR, $\psi$, and is parametrised as
\begin{equation}
\dot{M}_{\rm eject} = \beta(V_{\rm c})\,\psi = \left(\frac{V_{\rm c}}{V_{\rm SN}}\right)^{-\gamma_{\rm SN}}\psi{\rm,}
\end{equation}
where $V_{\rm c}$ is the circular velocity of the stellar disc/bulge at the half-mass radius and $V_{\rm SN}$ and $\gamma_{\rm SN}$ are adjustable parameters.  The metallicities of the cold gas and hot gas components are $Z_{\rm cold}$ and $Z_{\rm hot}$ respectively, and $\dot{M}_{\rm cool}$ is the rate at which gas cools from the hot to cold gas components.  The parameters in these equations (e.g. $V_{\rm SN}$ and $\gamma_{\rm SN}$) are then calibrated against a broad range of data from both observations and simulations, which places a strong constraint on the available parameter space, as described in L16.  

An exponential galactic disc is assumed to be formed by the cooling of hot halo gas, with a radius calculated assuming conservation of angular momentum and centrifugal equilibrium. Bulges/spheroids (with an assumed $r^{1/4}$ density profile) are formed by dynamical processes i.e. a disc instability or a galaxy merger, after which cold gas is moved from the disc into a newly formed bulge, where it is consumed in a starburst.  The size of the resulting bulge/spheroid is determined by virial equilibrium and energy conservation.  For more details regarding the computation of sizes in \galform, we refer the reader to Section 4.4 of \cite{Cole00}.      

Stellar luminosities are calculated using an evolutionary population synthesis model \citep[e.g.][]{BC03,Maraston05}, using the star formation and chemical enrichment histories predicted by \galform and using the assumed IMF.

Here we use as our fiducial model the version of \galform presented in L16, which incorporates a number of physical processes important for galaxy formation from earlier models.  These include a prescription for AGN feedback \citep{Bower06} in which quasi-hydrostatic hot halo gas is prevented from cooling by energy input from radiative jets, and a star formation law based on an empirical relation between molecular gas mass and star formation rate \citep{BlitzRosolowsky06}, first implemented in \galform by \cite{Lagos11}.  As previously mentioned, an important feature of the model is that it incorporates two modes of star formation: (i) a quiescent mode, in which star formation is fuelled by cold gas in the disc and in which stars form according to a solar neighbourhood \cite{Kennicutt83} IMF; and (ii) a starburst mode, in which star formation occurs in a bulge/spheroid formed by gas being transferred from the disc via a disc instability or a galaxy major merger (and some minor mergers) and in which stars form with a top-heavy IMF characterised by an unbroken $x=1$ slope in $\mathrm{d}N(m)/\mathrm{d}\ln m\propto m^{-x}$.  We will also present some results from the GP14 \galform model. The main difference between this model and the L16 model is that it assumes a universal \cite{Kennicutt83} IMF. However, unless otherwise stated results shown are for the L16 model.  For reference, the \cite{Kennicutt83} IMF used for quiescent mode star formation has a slope of $x=1.5$ (for $m>1$~M$_{\sun}$), whilst a \cite{Salpeter55} IMF has a slope of $x=1.35$.  These two modes of star formation are described in more detail below.

\subsubsection{Quiescent Mode}
In the quiescent mode the SFR is calculated according to the empirical \cite{BlitzRosolowsky06} relation, based on the fraction of molecular gas in the disc, $f_{\rm mol}$, which depends on the midplane gas pressure, $P$, at each radius in the disc  
\begin{equation}
R_{\rm mol} = \frac{\Sigma_{\rm mol}}{\Sigma_{\rm atom}} = \left[\frac{P}{P_{0}}\right]^{\alpha_{\rm P}}\rm ,
\end{equation}
where $R_{\rm mol}$ is the ratio of molecular to atomic gas at radius $r$, $\alpha_{\rm P}=0.8$ and $P_{0}/k_{\rm B}=1.7\times10^{4}$~cm$^{-3}$~K based on the observations of \cite{Leroy08}.  As mentioned earlier, it is assumed that the disc gas and stars are distributed in an exponential profile for the disc, the radial scalelength of which is predicted by \galform.  The star formation rate surface density, $\Sigma_{\rm SFR}$, at a given radius is then given by
\begin{equation}
\Sigma_{\rm SFR} = \nu_{\rm SF,quies}\,\Sigma_{\rm mol,disc}=\nu_{\rm SF,quies}\,f_{\rm mol}\Sigma_{\rm cold,disc}\rm,
\label{eq:quiescent_mode_SF}
\end{equation}
where $\Sigma_{\rm cold,disc}\propto\exp{(-r/h_{R})}$ ($h_{\rm R}$ is the radial scalelength), $f_{\rm mol} = R_{\rm mol}/(1+R_{\rm mol})$ and $\nu_{\rm SF,quies}=0.43$~Gyr$^{-1}$ \citep{Bigiel11}.  Equation (\ref{eq:quiescent_mode_SF}) is then integrated over the whole disc to derive the global star formation rate, $\psi$.  For further details regarding the prescription for quiescent mode star formation presented here see \cite{Lagos11}.  
\subsubsection{Burst Mode}
For star formation in bursts it is assumed that $f_{\rm mol}\approx1$ and the SFR depends on the dynamical timescale of the bulge
\begin{equation}
\psi_{\rm burst}= \nu_{\rm SF,burst}\,M_{\rm cold,burst}{\rm ,}
\end{equation}
where $\nu_{\rm SF,burst}=1/\tau_{\star\rm,burst}$ and
\begin{equation}
\tau_{\star\rm, burst}=\max[f_{\rm dyn}\tau_{\rm dyn,bulge},\tau_{\rm burst,min}]{\rm.}
\end{equation}
Here $\tau_{\rm dyn,bulge}=r_{\rm bulge}/V_{\rm c}(r_{\rm bulge})$, where $r_{\rm bulge}$ is the half-mass radius of the bulge, and $f_{\rm dyn}$ and $\tau_{\rm burst,min}$ are model parameters, such that for large dynamical times the SFR timescale scales with the dynamical timescale of the bulge ($f_{\rm dyn}=20$), but has a floor value ($\tau_{\rm burst,min}=100$~Myr) when the dynamical timescale of the bulge is short.

\subsubsection{Relation of inferred to true physical properties}
\label{subsubsec:infer}
One consequence of a model that has multiple IMFs is that it complicates the comparison of physical properties predicted by the model with those inferred from observations assuming a universal IMF.  Here, we scale the SFRs of the simulated galaxies to what would be inferred assuming a universal \cite{Kennicutt83} IMF. To do this we scale the burst mode SFR by a factor of $2.02$, assuming that infra-red luminosity is used as a tracer of star formation, as derived by GP14.  For stellar mass we use the SED fitting code presented in \cite{Mitchell13} to investigate if the top-heavy IMF in the model would have a significant impact on the inferred mass in Appendix~\ref{app:IMF}, and conclude that changes in the inferred stellar mass due to the top-heavy IMF are small and so we make no explicit correction for this here.  For the purposes of comparing to our model predictions  we also convert physical properties derived from various observational data to  what would have been inferred assuming a universal \cite{Kennicutt83} IMF, describing the conversion factors used in the text where relevant.  Throughout we denote these \emph{inferred} physical quantities by a prime symbol e.g. $M_{\star}^{\prime}$, sSFR$^{\prime}$.

\subsection{The Dust Emission Model}
\begin{table*}
\centering
\caption{List of quantities used in our dust model.  Those predicted by \galform\ are listed in the top part of the table.}
\begin{tabular}{lll}\hline
Quantity & Description & Value$^{\rm a}$ \\ \hline
\multicolumn{3}{l}{\galform quantities}\\
$M_{\rm cold}$&Mass of cold gas in the disc/burst &G \\
$Z_{\rm cold}$&Metallicity of cold gas in the disc/burst&G \\
$h_{\rm R}$&Radial scalelength of the disc/burst&G \\ \hline
\multicolumn{3}{l}{Fixed dust model parameters$^{\rm b}$}\\ 
$h_{z}/h_{R}$&Ratio of vertical to radial scalelengths&$0.1$ \\
$h_{z}\mathrm{(dust)}/h_{z}\mathrm{(stars)}$&Ratio of vertical scalelengths for dust and stars&$1$ \\
$\lambda_{\rm b}$&Break wavelength for dust opacity in bursts&$100$~$\muup$m\\
$m_{\rm cloud}/r_{\rm cloud}^{2}$&Molecular cloud parameters&$10^{6}$~M$_{\sun}$/($16$~pc)$^{2}$\\ \hline
\multicolumn{3}{l}{Adjustable dust model parameters} \\
$f_{\rm cloud}$&Fraction of gas/dust in molecular clouds&$0.5$ \\
$t_{\rm esc}$&Escape time of stars from molecular clouds&$1$~Myr\\
$\beta_{\rm b}$&Spectral index of dust opacity in bursts&$1.5$ \\\hline
\multicolumn{3}{l}{$^{\rm a}$A value of `G' indicates that this quantity is predicted by \galform.}\\
\multicolumn{3}{l}{$^{\rm b}$Variations of these parameters were not considered when calibrating the model (L16).}\\
\hline
\end{tabular}
\label{tab:dust_params}
\end{table*}  
To determine a simulated galaxy's FIR/sub-mm flux, a model is required to calculate the absorption and re-emission of its stellar radiation by interstellar dust.  Here we use a model in which the absorption is calculated by solving the equations of radiative transfer in an assumed geometry, and the dust temperature is calculated by solving for energy balance, assuming the dust emits as a modified blackbody.  Our model adopts a very similar geometry for the stars and dust to the spectrophotometric radiative transfer code, \grasil \citep{Silva98}.  However, for the purpose of reducing the computation time a number of simplifying assumptions are made relative to \grasil.  In this section we briefly describe our model, for further details see Appendix A of L16.

It is assumed that dust exists in two components: (i) dense molecular clouds of fixed gas surface density in which stars form, escaping on some timescale, $t_{\rm esc}$, such that stars begin to leave the cloud at time $t_{\rm esc}$ and they have all left after time $2t_{\rm esc}$ (the fiducial model uses a value of $t_{\rm esc}=1$~Myr); and (ii) a diffuse ISM which assumes the same scale lengths for the dust as for the stellar disc/burst.  The fraction of the dust mass in molecular clouds is determined by the parameter $f_{\rm cloud}$ (in the fiducial model $f_{\rm cloud}=0.5$).  The dust mass is calculated in \galform assuming a dust-to-gas ratio that scales linearly with cold gas metallicity, $Z_{\rm cold}$ (which is determined by equation \ref{eq:mzcolddot}), normalized to a local ISM value \citep[e.g.][]{Silva98} such that
\begin{equation}
M_{\rm dust} = \delta_{\rm dust}M_{\rm cold}Z_{\rm cold}\rm,
\label{eq:Mdust}
\end{equation}
where $\delta_{\rm dust}=0.334$.  This value for $\delta_{\rm dust}$ is derived assuming a value of $1/110$ for the dust-to-hydrogen gas mass ratio at solar metallicity \citep[e.g.][]{DraineLee84} and a fixed hydrogen-to-total gas mass ratio of $0.735$.  

The dust is assumed to have the same albedo, $a_{\lambda}$,  and extinction curve shape, $k_{\lambda}$, as in the solar neighbourhood \citep[e.g.][]{Silva98}.  The (extinction) optical depth for dust passing through some gas with surface density $\Sigma_{\rm gas}$ is
\begin{equation}
\tau_{\lambda\rm,ext}=0.043\left(\frac{k_{\lambda}}{k_{V}}\right)\left(\frac{\Sigma_{\rm gas}}{\mathrm{M}_{\sun}\,\mathrm{pc}^{-2}}\right)\left(\frac{Z_{\rm cold}}{0.02_{}}\right)\rm,
\label{eq:central_optical_depth}
\end{equation} where the normalisation is based on the local ratio of $V$-band extinction to hydrogen column density, as measured by \cite{SavageMathis79}.  For example, for diffuse disc gas $\Sigma_{\rm gas}=\left(1-f_{\rm cloud}\vphantom{\frac{}{}}\right)\,\left(M_{\rm cold,disc}/2\pi\,h_{\rm R}^{2}\right)\exp\left(-r/h_{R\rm, disc}\vphantom{\frac{}{}}\right)$ where $h_{R\rm, disc}$ is the radial scalelength of the disc, as predicted by \galform; for molecular clouds the gas surface density through the centre of a cloud is $\Sigma_{\rm gas}=3m_{\rm cloud}/4\pi\,r_{\rm cloud}^{2}$.

In molecular clouds the effective absorption optical depth is approximated as
\begin{equation}
\tau_{\lambda\rm,eff}=(1-a_{\lambda})^{1/2}\tau_{\lambda\rm,ext}\rm,
\end{equation}from which the dust attenuation due to clouds averaged over stellar age, $t$, can be computed as
\begin{equation}
\langle A_{\lambda}^{\rm MC}\rangle = 1 - \langle\eta(t)\rangle(e^{-\tau_{\lambda\rm,eff}}-1)\rm.
\end{equation} Here $\eta(t)$ is the fraction of stars of age $t$ that are still in their birth clouds, and is parametrised as
\begin{equation}
\eta(t)=
\begin{cases}
1 & t<t_{\rm esc}\\
2-t/t_{\rm esc} & t_{\rm esc}<t<2t_{\rm esc}\\
0 & t>2t_{\rm esc}\rm,
\end{cases}
\end{equation} as described above and introduced in \cite{Silva98}, to model the complex process of molecular cloud disruption.

For the diffuse component the attenuation factor, $A_{\lambda}^{\rm diff}$, is calculated from the tabulated radiative transfer results of \cite{Ferrara99}, which are computed using the code described in \cite{BFG96}.  For this purpose it is assumed that the ratio of vertical to radial scalelengths of the disc is $h_{z}/h_{R}=0.1$, and that the dust and stars have the same scaleheight i.e. $h_{z}(\mathrm{dust})=h_{z}(\mathrm{stars})$.  These assumptions are discussed in more detail in \cite{Granato00}.  The central optical depth is calculated according to equation (\ref{eq:central_optical_depth}) and the Ferrara et al. tables are interpolated to get the total attenuation as a function of wavelength, $A^{\rm diff}_{\lambda}$.  If a galaxy is undergoing a starburst a similar procedure is followed but using the gas mass and radial scalelength of the forming bulge/spheroid (burst), which are predicted by \galform. 

With the dust attenuation factors for each dust component it is possible to compute the energy they each absorb.  For the molecular clouds this is given by
\begin{equation}
L_{\rm abs}^{\rm MC}=\int_{0}^{\infty}\left(1-\left\langle A_{\lambda}^{\rm MC}\right\rangle\right)L_{\lambda}^{\rm unatten}\mathrm{d}\lambda\rm,
\end{equation}where $L_{\lambda}^{\rm unatten}$ is the unattenuated stellar SED of the galaxy predicted by \galform.  For the diffuse dust, the energy absorbed is given by
\begin{equation}
L_{\rm abs}^{\rm diff}=\int_{0}^{\infty}\left(1-A_{\lambda}^{\rm diff}\right)\,\left\langle\,A_{\lambda}^{\rm MC}\right\rangle\,L_{\lambda}^{\rm unatten}\,\mathrm{d}\lambda\rm.
\end{equation} 

Assuming thermal equilibrium, and that the dust is optically thin to its own emission, the energy absorbed by each component is equated to the emission from a modified blackbody,
\begin{equation}
L_{\lambda}^{\rm dust} = 4\pi\,\kappa_{\rm d}(\lambda)\,B_{\lambda}(T_{\rm d})\,Z_{\rm cold}\,M_{\rm cold}\rm,
\label{eq:galform_dust_emission}
\end{equation}         
here $T_{\rm d}$ is the dust temperature, $B_{\lambda}(T_{\rm d})$ is the Planck blackbody function and $\kappa_{\rm d}(\lambda)$ is the dust opacity per unit mass of metals in the gaseous phase.  This is parametrised as
\begin{equation}
\label{eq:kappa_d}
\kappa_{\rm d}(\lambda)=
\begin{cases}
\kappa_{1}\left(\frac{\lambda}{\lambda_{1}}\right)^{-2}&\lambda<\lambda_{\rm b}\\
\kappa_{1}\left(\frac{\lambda_{\rm b}}{\lambda_{1}}\right)^{-2}\left(\frac{\lambda}{\lambda_{\rm b}}\right)^{-\beta_{\rm b}}&\lambda>\lambda_{\rm b}\mathrm{,}\\
\end{cases}
\end{equation}
with $\kappa_{1}=140$~cm$^2$g$^{-1}$ at a reference wavelength of $\lambda_{1}=30$~$\muup$m \citep{DraineLee84}.  For burst mode star formation $\lambda_{\rm b}=100$~$\muup$m and for quiescent mode star formation an unbroken power law is assumed, equivalent to $\lambda_{\rm b}\rightarrow\infty$.  A value of $\beta_{\rm b}=1.5$ is used in the fiducial model.  This value for $\beta_{\rm b}$ is compatible with laboratory measurements \citep[e.g.][]{Agladze96} which suggest that values in the range $\beta_{\rm d}=1.5-2$ are acceptable.  Additionally, more recent observational measurements suggest an anti-correlation between $\beta_{\rm d}$ and temperature, with hotter dust exhibiting lower values of $\beta_{\rm d}$ (Boudet et al. \citeyear{Boudet05}).  Again this would be compatbile with our model as starburst galaxies generally exhibit hotter dust temeratures than quiescent galaxies.  The total FIR SED for each galaxy is then calculated by simply summing the SEDs of each dust component.

We emphasise that this means that the dust temperature is not a parameter of the model, but that it is calculated self-consistently given the assumptions regarding global energy balance and that each dust component is described by a single temperature.  We have listed the quantities used by our dust model in Table~\ref{tab:dust_params}. 

Despite the simplicity of this model, comparisons have shown that it can accurately reproduce the results of a more complete radiative transfer calculation performed by using the spectrophotometric code \grasil \citep{Silva98} for rest-frame wavelengths of $\lambda_{\rm rest}\gtrsim70$~$\muup$m.  This is where the emission is dominated by cool ($T_{\rm d}\sim20-40$~K) dust in thermal equilibrium. We illustrate this in Appendix \ref{app:grasil_comparison} by comparing the luminosities predicted by \grasil to those predicted by the simple dust model described here.  We indicate throughout the paper the wavelength range over which the simplifying approximations made in the simple dust model (e.g. each dust component described by a single temperature, dust is optically thin to its own emission) are valid.    

\section{Results}
\label{sec:Results}
Here we present our main results. In Section~\ref{subsec:sSFR_Mstar_plane} we show model predictions for the distribution of galaxies on the specific star formation rate (sSFR$^{\prime}$)~$-$~$M_{\star}^{\prime}$ plane [where the prime symbol indicates that these properties have been scaled to what would be inferred assuming a universal Kennicutt (\citeyear{Kennicutt83}) IMF as described in Section~\ref{subsubsec:infer}], describe our identification of a main sequence of star-forming galaxies, and how we define samples of galaxies selected based on their position relative to this main sequence.  In Section~\ref{subsec:stacked_SEDs} we then discuss the stacked SEDs of MS and SB galaxies selected in this way, and the trends we find in dust mass ($M_{\rm dust}$) and total FIR luminosity ($L_{\rm IR}$).  In Section~\ref{subsec:Compare} we perform a detailed comparison to the observations presented in \cite{Bethermin15}, and we investigate if these can provide any further constraints on the parameters of our dust model in Section~\ref{subsec:dust_params}.  
\subsection{The specific star formation rate - stellar mass plane}
\label{subsec:sSFR_Mstar_plane}
\begin{figure}
\includegraphics[width=0.99\linewidth]{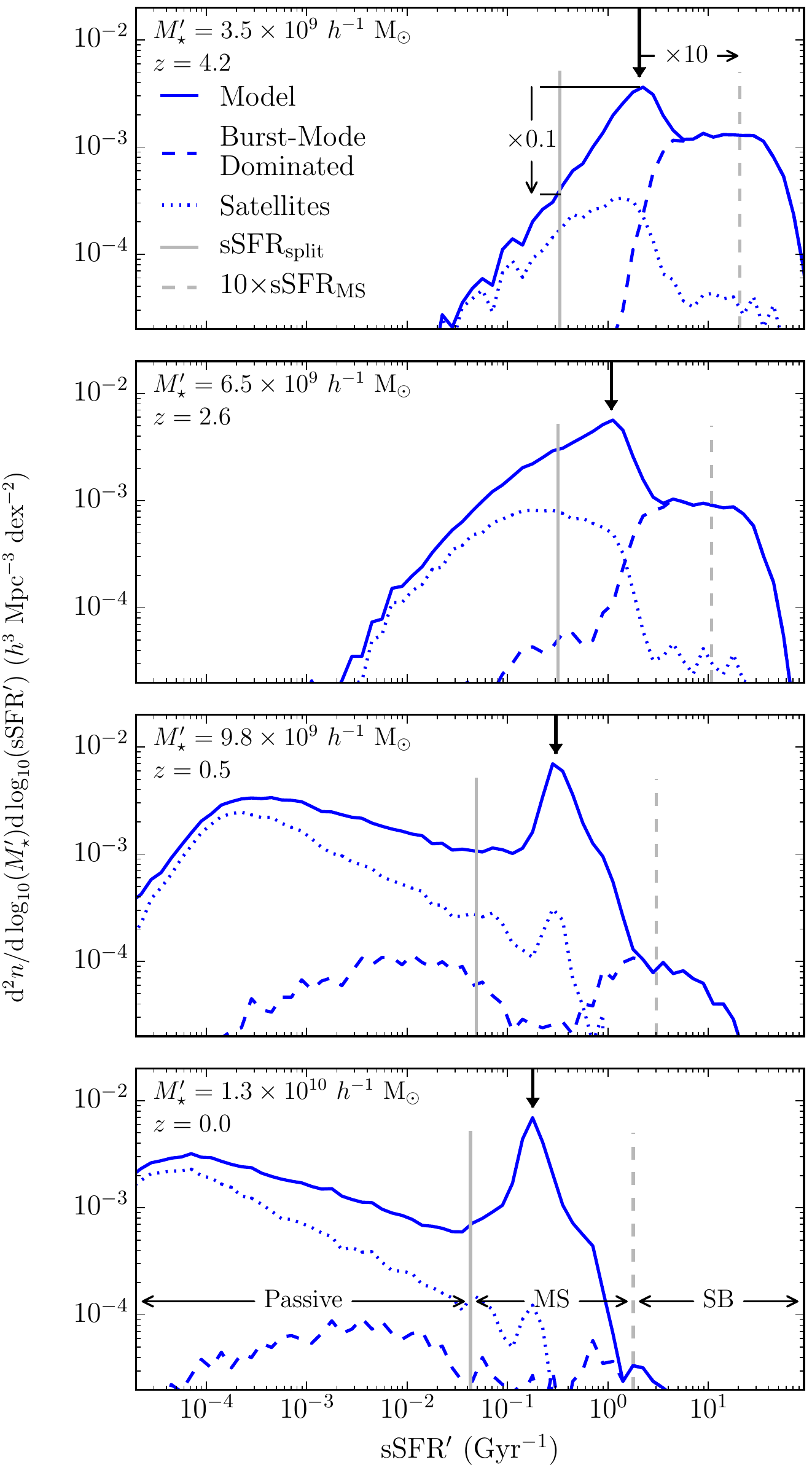}
\caption{Inferred specific star formation rate [sSFR$^{\prime}$, where the prime indicates the value inferred assuming a universal Kennicutt (\citeyear{Kennicutt83}) IMF] distributions in various stellar mass mass bins ($0.2$~dex width) at redshifts  $z=4.2$, $2.6$, $0.5$ and $0.0$ (top to bottom panels respectively).  The centre of the stellar mass bin is the best fitting value for $M_{\rm bk}$ in equation~(\ref{eq:doubleSchechter}) at the redshift indicated in the panel.  The dashed and dotted lines show the contribution to the total inferred sSFR distribution for burst mode dominated galaxies and satellite galaxies respectively. The thick black downward arrow indicates the position of sSFR$_{\rm peak}^{\prime}$.  By construction this is equal to sSFR$_{\rm MS}^{\prime}(M_{\star},z)$ at this stellar mass and redshift (as $M_{\star}^{\prime} = M_{\rm bk}$).  The vertical grey solid line indicates the split between star-forming and passive galaxies (sSFR$_{\rm split}^{\prime}$) and the vertical grey dashed line indicates the split between main sequence (MS) and starburst (SB) populations (i.e. $f_{\rm SB}\times\mathrm{sSFR}_{\rm MS}^{\prime}$, here $f_{\rm SB}=10$).}
\label{fig:Npsi}
\end{figure}
\begin{figure*}
\includegraphics[trim = 0 0 0 0,clip = true,width=\linewidth]{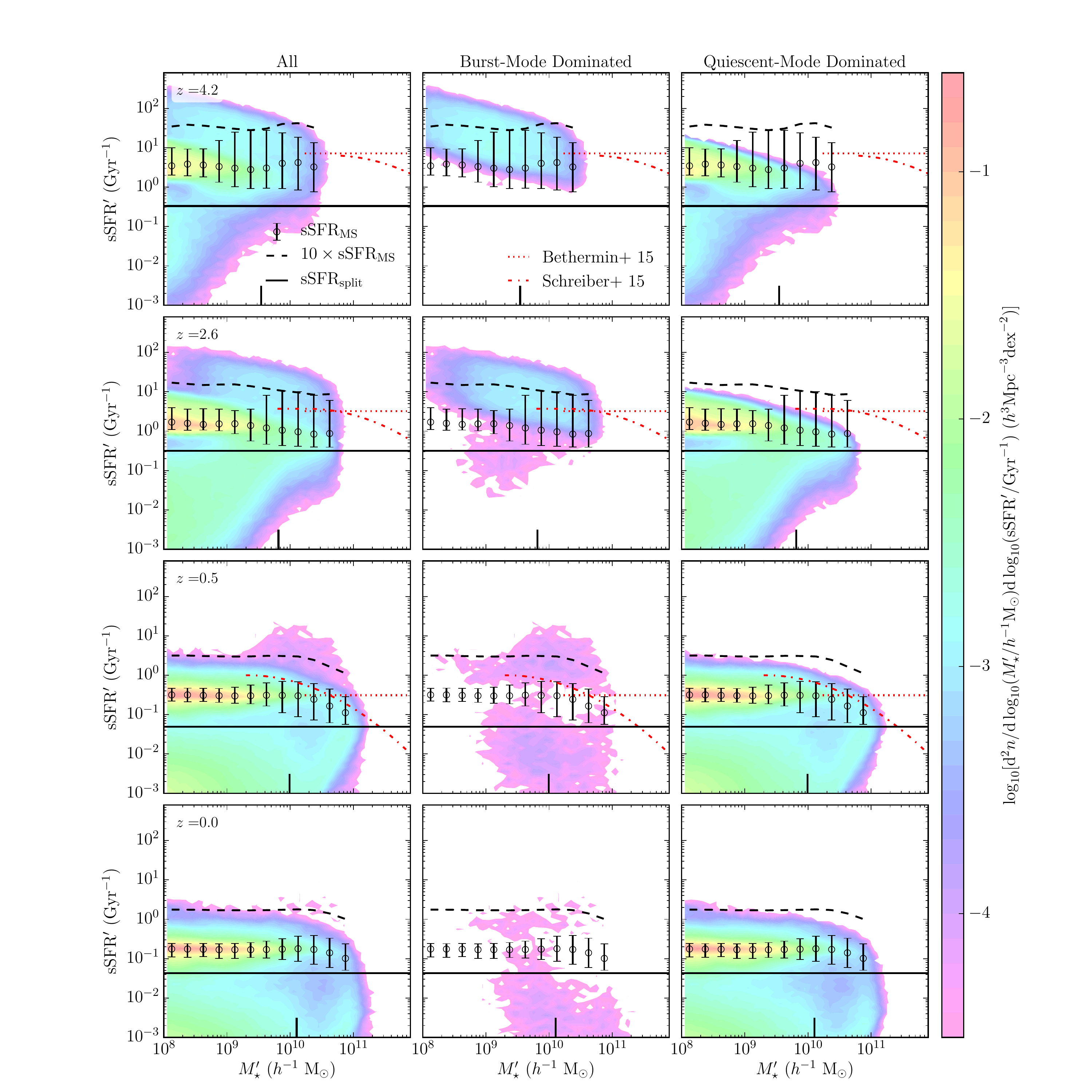}
\caption{The predicted galaxy number density in the sSFR$^{\prime}$-$M_{\star}^{\prime}$ plane at redshifts $z=4.2$, $2.6$, $0.5$ and $0.0$ (top to bottom rows respectively) for all galaxies (left panels), for burst mode dominated galaxies (middle panels) and for quiescent mode dominated galaxies (right panels).  The prime indicates that the value for that property is what would be inferred assuming a universal Kennnicutt (\citeyear{Kennicutt83}) IMF.  The colour scale indicates the predicted density of galaxies on this plane as shown in the key on the right.  The horizontal black line indicates sSFR$^{\prime}_{\mathrm{split}}$, above which galaxies are defined as star-forming.  The open circles show the median sSFR of star-forming galaxies in logarithmic stellar mass bins i.e. sSFR$^{\prime}_{\rm{MS}}(M_{\star},z)$, whilst the errorbars show the $16-84$ ($1\sigma$) percentile ranges of star-forming galaxies.  The black dashed line is $f_{\rm SB}\times$sSFR$^{\prime}_{\rm MS}$, here $f_{\rm SB}=10$.  Galaxies that lie above this line are defined as SB galaxies.  The heavy vertical black tick mark is the characteristic stellar mass ($M_{\rm bk}$) at that redshift, derived from fitting equation~(\ref{eq:doubleSchechter}) to the predicted galaxy stellar mass function.  The red dotted and dash-dotted lines are the observational estimates of the position of the star-forming main sequence from Schreiber et al. (\citeyear{Schreiber15}) and Bethermin et al (\citeyear{Bethermin15}) respectively, scaled to a Kennicutt (\citeyear{Kennicutt83}) IMF as described in the text, over the ranges of redshift and inferred stellar mass for which these estimates are valid.}
\label{fig:sSFR_SM_Planes}
\end{figure*}
\begin{figure}
\centering
\includegraphics[width=\linewidth]{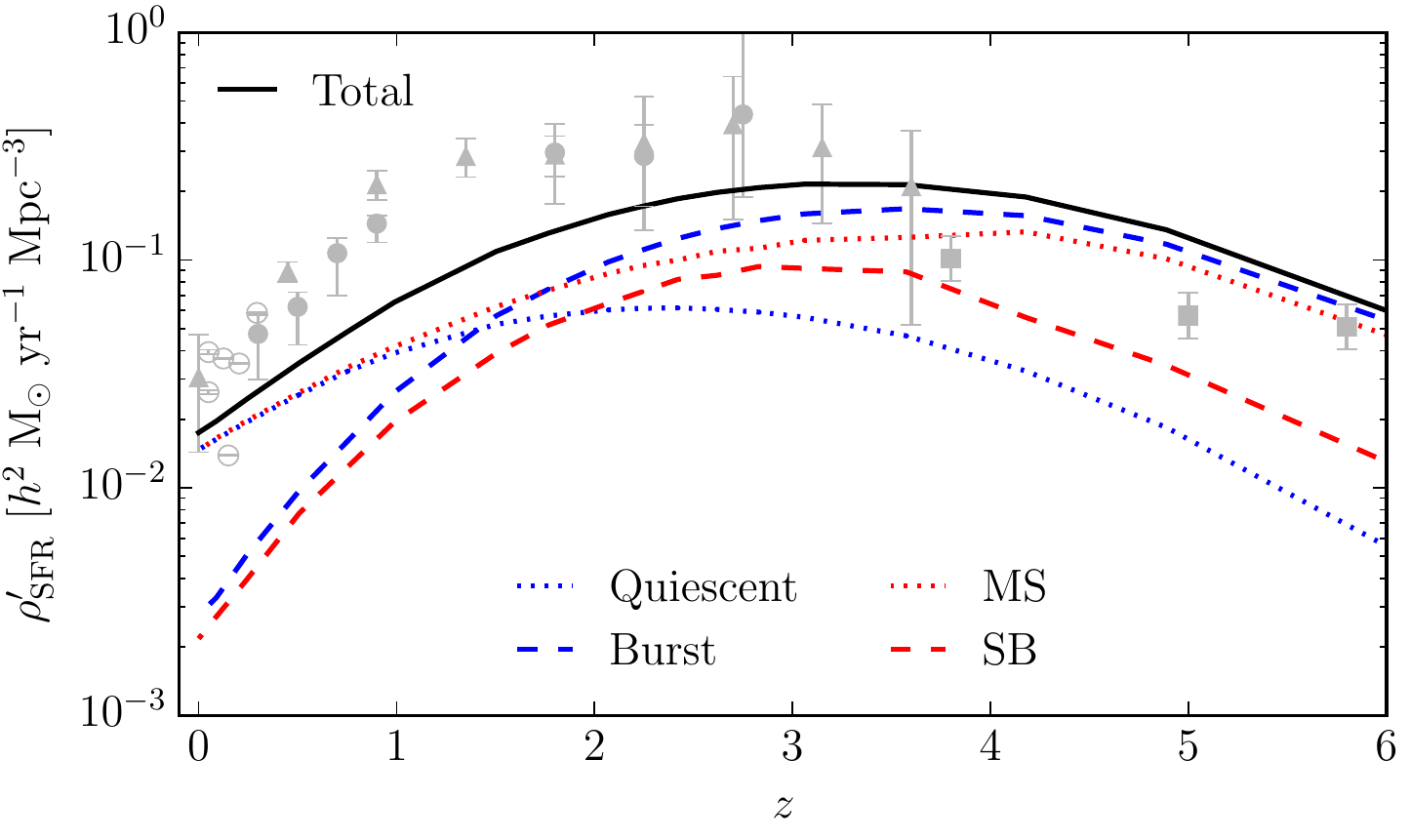}
\caption{The predicted comoving SFR density as a function of redshift [as inferred assuming a universal Kennicutt (\citeyear{Kennicutt83}) IMF, see Section~\ref{subsubsec:infer}].  The black line shows the total.  The blue dashed and dotted lines show the contribution from galaxies which are respectively burst mode dominated and quiescent mode dominated in the model.  The red dashed and dotted lines show the contribution from SB and MS galaxies respectively, classified according to their position on the sSFR$^{\prime}$-$M_{\star}^{\prime}$ plane.  Observational data (grey points with errorbars) are from Burgarella et al. (\citeyear{Burgarella13}, triangles), Gunawardhana et al. (\citeyear{Gunawardhana13}, open circles), Oesch et al. (\citeyear{Oesch12}, squares) and Karim et al. (\citeyear{Karim11}, filled circles).  Observational data have been scaled to what would be inferred assuming a Kennicutt (\citeyear{Kennicutt83}) IMF using conversion factors derived by GP14 and L16.}
\label{fig:rho_SFR}
\end{figure}
\begin{figure}
\centering
\includegraphics[width=\linewidth]{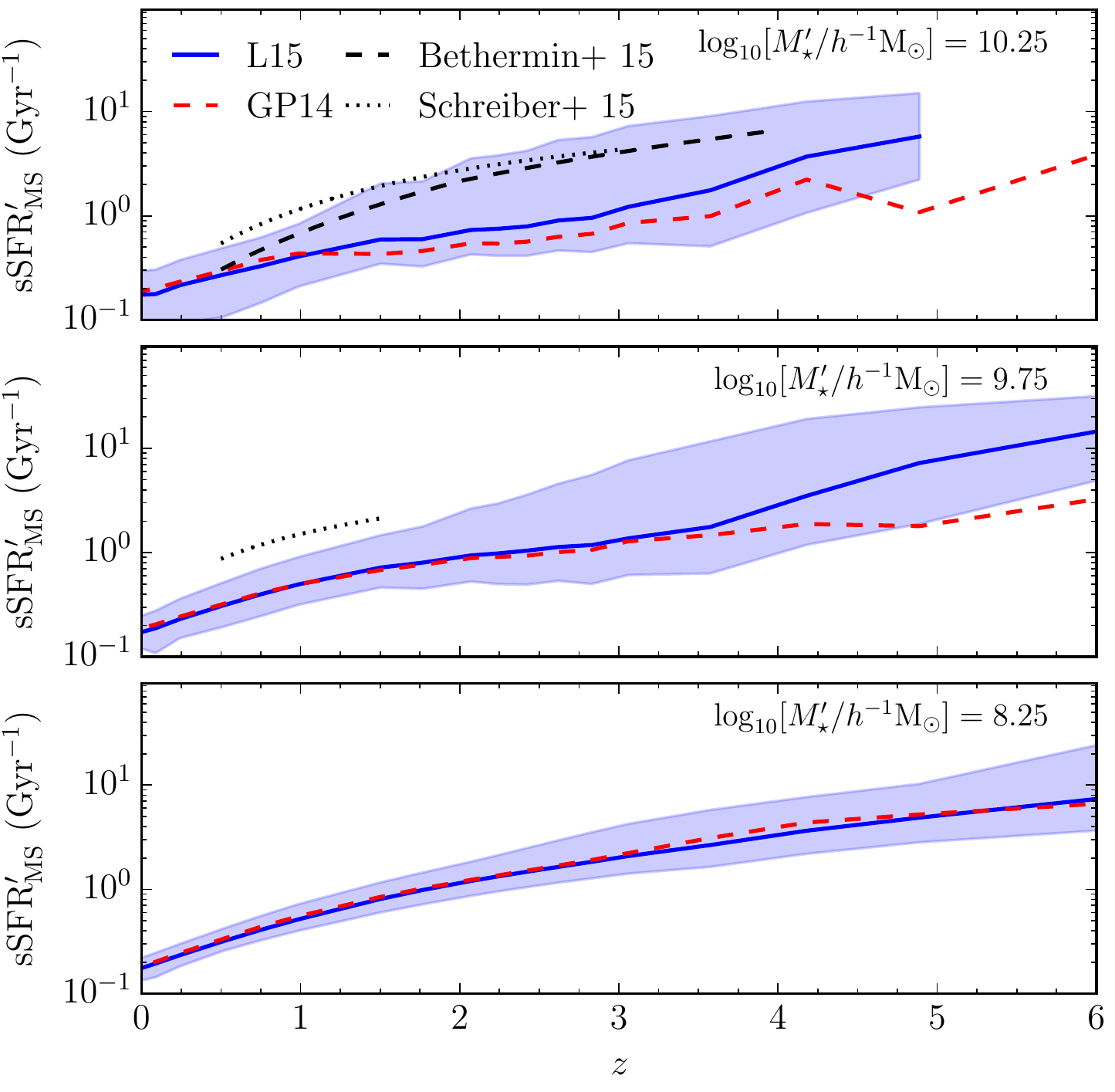}
\caption{Redshift evolution of sSFR$^{\prime}_{\rm MS}$ at fixed stellar masses indicated in the panels. In each panel the same stellar mass is used at all redshifts.  Predictions are shown from the L16 (blue solid line) and GP14 (red dashed line) models.  The blue shaded region indicates the $16-84$ ($1\sigma$) percentile scatter around sSFR$^{\prime}_{\rm MS}$ at that redshift and stellar mass for the L16 model. The black dashed and dotted lines show respectively observational estimates from Bethermin et al. (\citeyear{Bethermin15}) and Schreiber et al. (\citeyear{Schreiber15}), scaled to a Kennicutt (\citeyear{Kennicutt83}) IMF as described in the text, over the range in stellar mass and redshift for which these estimates are valid.}
\label{fig:sSFR_MS_z}
\end{figure}
As we are concerned with galaxies selected by their location on the sSFR$^{\prime}$-$M_{\star}^{\prime}$ plane, we define a redshift-dependent sSFR$_{\rm split}^{\prime}$, which separates star-forming and passive galaxies.  To do this we fit a double Schechter function with a single value for the mass break ($M_{\rm bk}$)
\begin{equation}
\phi(M_{\star}^{\prime})\,\mathrm{d}M_{\star}^{\prime}=e^{-M_{\star}^{\prime}/M_{\mathrm{bk}}}\left[\phi_{1}\left(\frac{M_{\star}^{\prime}}{M_{\rm bk}}\right)^{\alpha_{1}}+\phi_{2}\left(\frac{M_{\star}^{\prime}}{M_{\rm bk}}\right)^{\alpha_{2}}\right]\frac{\mathrm{d}M_{\star}^{\prime}}{M_{\star}^{\prime}}\rm,
\label{eq:doubleSchechter}
\end{equation}
(e.g. Baldry et al. \citeyear{Baldry12}) to the galaxy stellar mass function.  This provides a best-fit characteristic stellar mass, $M_{\rm bk}$, at each output redshift.  We then investigate the sSFR$^{\prime}$ distribution at this stellar mass ($\pm0.1$~dex), identifying a well defined peak (at sSFR$_{\rm peak}^{\prime}$) at high inferred sSFRs ($10^{-2}$~$<$~$\psi^{\prime}/M_{\star}^{\prime}$~$<$~$10$~Gyr$^{-1}$).  The value of sSFR$_{\rm split}^{\prime}$, indicated by the vertical solid lines in Fig.~\ref{fig:Npsi}, is then chosen so that by construction, at this characteristic mass, the median inferred sSFR for all galaxies with sSFR$^{\prime}>$sSFR$^{\prime}_{\rm split}$ is equal to sSFR$^{\prime}_{\rm peak}$.  In cases where this is not well defined i.e. sSFR$_{\rm peak}^{\prime}$ is less than the median inferred sSFR for all galaxies at that redshift, we simply set sSFR$^{\prime}_{\rm split}$ to be the inferred sSFR at which the distribution, $\mathrm{d}n/\mathrm{d}\log_{10}\mathrm{sSFR}^{\prime}$, is equal to a tenth of its maximum value (for sSFRs$^{\prime}<$sSFR$^{\prime}_{\rm peak}$, see the top panel of Fig.~\ref{fig:Npsi} for an example of this).  In this manner we have a well defined method for choosing sSFR$_{\rm split}^{\prime}$ at each redshift that is not dependent on observations or on choosing sSFR$_{\rm split}^{\prime}$ by eye.  We prefer this method to using rest-frame near-UV/optical colours to separate passive and star-forming galaxies, as this would be overly sensitive to assumptions made in the model about the details of the dust attenuation.  We use a single sSFR$_{\rm split}^{\prime}$ at each redshift (i.e. independent of inferred stellar mass) for simplicity. We do not expect this to significantly affect our results as this assumption has a minor impact on the position of the main sequence.     

The inferred sSFR of the main sequence of star-forming galaxies, which depends on stellar mass as well as redshift, sSFR$_{\rm MS}^{\prime}(M_{\star},z)$, is then defined as the median inferred sSFR for all galaxies with sSFRs$^{\prime}>$sSFR$^{\prime}_{\rm split}$ at a given inferred stellar mass and redshift.  We define galaxies as main sequence (MS) if they have sSFR$^{\prime}_{\rm split}<\mathrm{sSFR}^{\prime}<f_{\rm SB}\times\mathrm{sSFR}^{\prime}_{\rm MS}$,  as starbursts (SB) if they have ${\rm sSFR}^{\prime}>f_{\rm SB}\times\mathrm{sSFR}^{\prime}_{\rm MS}$, and as passive if they have sSFR$^{\prime}<$sSFR$^{\prime}_{\rm split}$.  This demarcation is shown in the bottom panel of Fig.~\ref{fig:Npsi}.  We use $f_{\rm SB}=10$ throughout to distinguish SB and MS galaxies. This choice is somewhat arbitrary but motivated by the value used in observational studies \citep[e.g.][]{Bethermin15}.

We can see in Fig.~\ref{fig:Npsi} that the passive galaxy population is dominated by satellite galaxies.  The star formation in these galaxies is inhibited by diminishing cold gas reservoirs. In our model a galaxy's hot gas halo is removed by instantaneous ram-pressure stripping upon becoming a satellite and it is assumed that no further gas will cool onto it [see \cite{Lagos14b} for an analysis of the effect this modelling has on the atomic and molecular gas content of galaxies].      

Now that we have defined our galaxy populations, in Fig.~\ref{fig:sSFR_SM_Planes} we show the predicted distribution of galaxies on the sSFR$^{\prime}$-$M_{\star}^{\prime}$ plane at a range of redshifts, and separated by the mode of star formation.  We note that the definition of SB which uses a galaxy's position on the sSFR$^{\prime}$-$M_{\star}^{\prime}$ plane, is not the same as a model galaxy being dominated by burst mode star formation.  In the middle panels of Fig.~\ref{fig:sSFR_SM_Planes} we can see that these two definitions are somewhat different, and that many galaxies dominated by burst mode star formation would be classified as MS based on their position on the sSFR$^{\prime}$-$M^{\prime}_{\star}$ plane.  We further emphasize this point in Fig.~\ref{fig:rho_SFR}, where we show the contribution to the total comoving inferred SFR density predicted by the model for the MS and SB samples (red dotted and red dashed lines respectively) and for galaxies dominated by quiescent and burst mode star formation (blue dotted and blue dashed lines respectively).  We can see here that whilst the MS sample dominates the inferred SFR density, contributing $\sim65$~per~cent at all redshifts, at higher redshifts ($z\gtrsim1.5$) it is burst mode dominated galaxies that make the dominant contribution to the inferred SFR density.  We note that the precise contribution of the MS to the inferred star formation rate density is somewhat sensitive to our definition of MS.  If we reduce the value of $f_{\rm SB}$ to $4$ then the MS contribution to the total drops to $\sim50$~per~cent.  

We also note that the population of `passive bursts' (i.e. burst mode dominated galaxies that lie below the MS) evident in the panels of the middle column of Fig.~\ref{fig:sSFR_SM_Planes} comprises galaxies in which burst mode star formation was triggered by a galaxy merger.  The main locus of burst mode dominated galaxies is populated by disc instability triggered bursts.  

Our method of defining sSFR$^{\prime}_{\rm MS}(M_{\star},z)$ allows us to investigate the scatter around this relation.  We can see from the errorbars shown in Fig.~\ref{fig:sSFR_SM_Planes}, which indicate the $16-84$ percentile ($1\sigma$) scatter of star-forming galaxies around sSFR$^{\prime}_{\rm MS}(M_{\star},z)$, that the scatter tends to be smaller at lower stellar masses and at lower redshifts.  We can understand this in terms of how a galaxy regulates its star formation.  At low stellar masses ($M_{\star}^{\prime}\lesssim10^{10}$~$h^{-1}$~M$_{\sun}$) and in quiescent mode a galaxy's gas supply (and hence star formation) is self-regulated through the interplay of accretion of matter onto the halo, and the prescriptions for gas cooling from the hot halo and stellar feedback in the model, which produces a tight relationship between the SFR and stellar mass of a galaxy \citep[][see also Mitchell et al. \citeyear{Mitchell14}]{Lagos11}.  

When this is not the case the relationship between SFR and stellar mass becomes weaker, resulting in a larger scatter. This can be due to a number of reasons which we now discuss in turn: 

(i) \emph{Burst mode star formation} in which star formation is enhanced due to some dynamical process. The sSFR distributions of burst-mode dominated galaxies tend to be broader (Fig.~\ref{fig:Npsi}), this has more of an effect at higher redshifts where the burst mode contributes more to the global star formation density. 

(ii) \emph{Environmental effects} such as ram-pressure stripping restricting a galaxy's gas supply.  This affects satellite galaxies in our model and is the reason why they form the bulk of our passive galaxy population.

(iii) \emph{AGN feedback} in massive ($M_{\rm h}\gtrsim10^{12}$~$h^{-1}$~M$_{\sun}$) halos, which generally affects galaxies with $M_{\star}^{\prime}\gtrsim10^{10}$~$h^{-1}$~M$_{\sun}$. Whilst increasing the scatter, this physical process also inhibits star formation, giving rise to the negative slope seen in the sSFR$^{\prime}_{\rm MS}$ at high stellar masses in the bottom two rows of Fig.~\ref{fig:sSFR_SM_Planes}.  This negative slope at high stellar masses is also reflected in these galaxies being bulge dominated \citep[e.g.][]{Abramson14,Schreiber16}. For example, in our model at $z=0$ we find that MS galaxies with $M_{\star}^{\prime}>10^{10}$~$h^{-1}$~M$_{\sun}$ have a median bulge-to-total ratio of stellar mass of $B/T=0.5$, whereas galaxies with lower stellar masses ($10^{9}<M_{\star}^{\prime}<10^{10}$~$h^{-1}$~M$_{\sun}$) have a median ratio of $B/T=0.002$.  A galaxy's bulge and supermassive black hole are grown by the same processes in the model (disc instability or galaxy merger) and so it is not surprising that they are linked. This also evidenced in the anti-correlation between cold gas fraction and bulge mass found by \cite{Lagos14b}. Galaxies with larger bulges are likely to have more massive SMBHs, and therefore more effective AGN feedback that inhibits gas cooling, leading to suppressed star formation rates.

As discussed earlier, we have scaled the observational estimates of the position of the main sequence from \cite{Schreiber15} and \cite{Bethermin15} that appear in Fig.~\ref{fig:sSFR_SM_Planes} to what would be inferred assuming a universal \cite{Kennicutt83} IMF.  We scale the \cite{Schreiber15} SFRs, which were derived assuming a \cite{Salpeter55} IMF and using UV + IR as a tracer for star formation by a factor of $0.8$. The \cite{Bethermin15} SFRs, derived using $L_{\rm IR}$ as a tracer for star formation and assuming a \cite{Chabrier03} IMF are scaled by a factor of $1.29$.  These conversion factors were calculated by GP14, using the PEGASE.2 SPS model \citep{Fioc97}.  For stellar masses we scale the Schreiber et al. masses by a factor of $0.47$ \citep[Salpeter to Kennicutt IMF, ][]{Ilbert10} and the B\'{e}thermin et al. masses by $0.81$ \citep[Chabrier to Kennicutt, ][]{Santini12}.  The mass limit of the B\'{e}thermin et al. sample, quoted as $3\times10^{10}$~M$_{\sun}$, becomes $1.7\times10^{10}$~$h^{-1}$~M$_{\sun}$, also accounting for the factor of $h=0.7$ assumed by those authors.

Also evident in Fig.~\ref{fig:sSFR_SM_Planes} is the global trend of increasing sSFR$^{\prime}_{\rm MS}$ with redshift.  We show the redshift evolution of sSFR$^{\prime}_{\rm MS}$, and its $1\sigma$ percentile scatter, at a range of fixed stellar masses for both the L16 (blue line) and GP14 (red dashed line) models in Fig.~\ref{fig:sSFR_MS_z}.  The models agree qualitatively with the observational data insofar as they both predict an increasing sSFR$^{\prime}_{\rm MS}$ with increasing redshift.  However, the predicted normalisation does not agree with the observed value at all stellar masses and redshifts.  For example, the models appear to underpredict the sSFR$^{\prime}_{\rm MS}$ for $0.5\lesssim z\lesssim4$ for $M_{\star}^{\prime}\gtrsim10^{10}$~$h^{-1}$~M$_{\sun}$ by a factor of $\sim2$.  It is worth noting that both observational studies use the relation of \cite{Kennicutt98} to convert from observed $L_{\rm IR}$ to inferred SFR.  This relation was derived initially for dusty circumnuclear starbursts (for a burst duration of $\lesssim100$~Myr) in which the total bolometric luminosity of the stellar population is assumed to be re-radiated in the infra-red, an assumption that may not be wholly valid for the main sequence galaxies considered here.  In addition, the evolution of the main sequence at low redshift ($z\lesssim2$) is not as strong as implied by the observations.  This is similar to what was found in a study performed using an earlier version of the \galform models used here by \cite{Mitchell14}, who used the model of \cite{Lagos12} but with the continuous gas cooling model proposed by Benson \& Bower (\citeyear{BensonBower10}).  Mitchell et al. attributed this discrepancy to the stellar mass assembly histories of the galaxies predicted by \galform being approximately flat for $z\lesssim2$, driven primarily by the level of coevolution between stellar and dark matter halo mass assembly in the model, whereas the stellar mass assembly history inferred from observations decreases over the same epoch.  Both \galform models shown here predict very similar evolution for sSFR$^{\prime}_{\rm MS}$, differing only at high masses and redshifts. This happens where the contribution to the MS from burst-mode dominated galaxies is most significant, with the top-heavy IMF allowing the L16 model to have generally a higher sSFR$^{\prime}_{\rm MS}$ after adjusting to a universal Kennicutt (\citeyear{Kennicutt83}) IMF.  

The $16-84$ ($1\sigma$) percentile scatter around the main sequence for the L16 model is shown in Fig.~\ref{fig:sSFR_MS_z} as the shaded blue region.  At $z=1$ this is $0.26$, $0.5$ and $0.6$~dex respectively for $M_{\star}^{\prime}=10^{8.25}$, $10^{9.25}$ and $10^{10.25}$~$h^{-1}$~M$_{\sun}$, with the scatter being smaller for lower stellar masses as this is where the main sequence is dominated by quiescent mode star formation.  This scatter is approximately constant for $z\lesssim1.5$ for the two higher stellar masses and for $z\lesssim2$ for the lowest, increasing at higher redshift.  These results are in qualitative agreement with the findings of \cite{Ilbert15}, who find that scatter around the MS increases with stellar mass and is independent of redshift up to at least $z\sim1.4$.  
 
\subsection{Stacked infra-red SEDs}
\label{subsec:stacked_SEDs}
\begin{figure}
\centering
\includegraphics[trim = 0 10 0 0,clip = True,width=\linewidth]{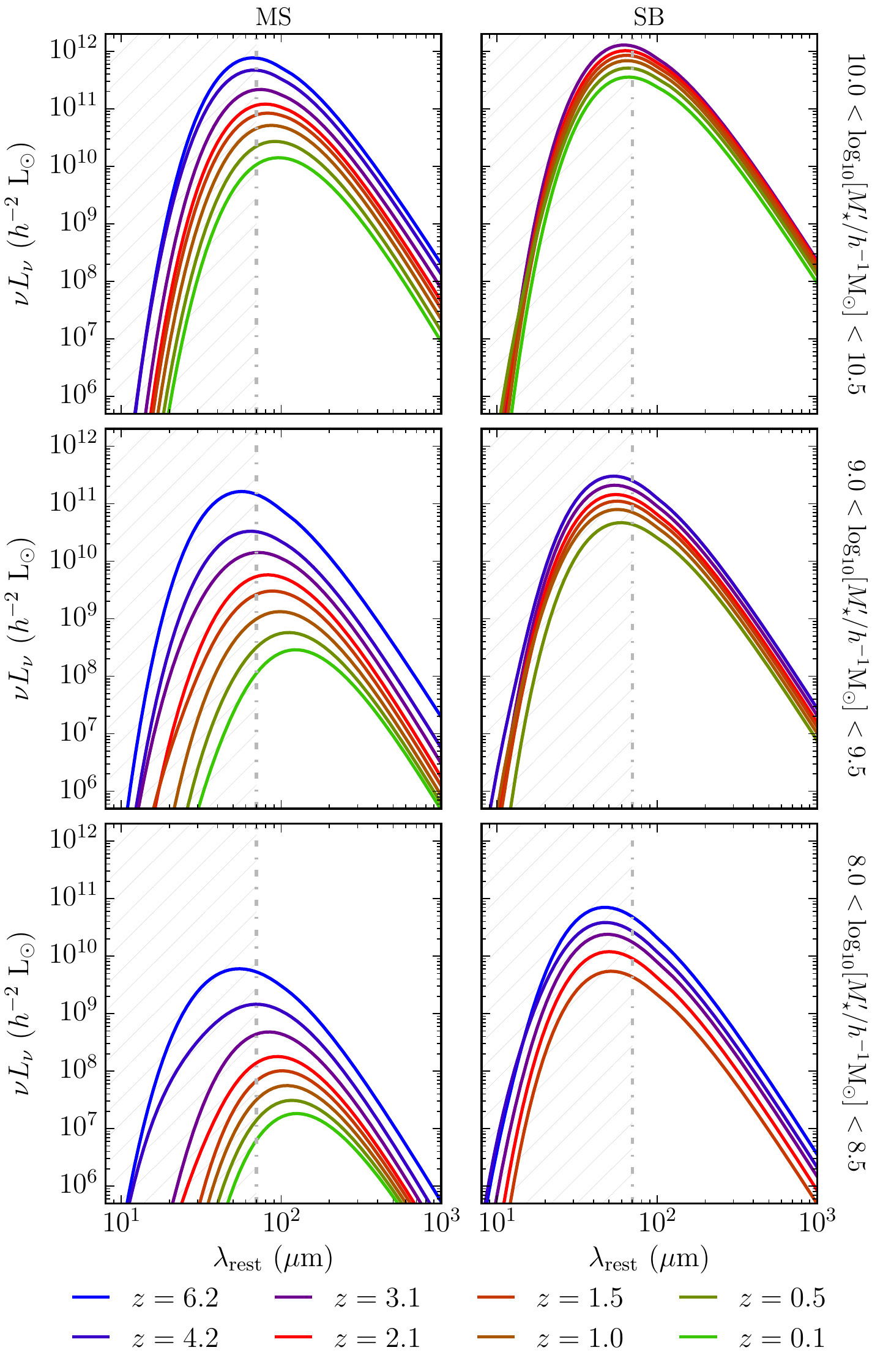}
\caption{Redshift evolution of stacked SEDs for main sequence (left column) and starburst galaxies (right column), for galaxies selected by their stellar mass at the redshift in question, for $10<\log_{10}[M_{\star}/h^{-1}~\mathrm{M}_{\sun}^{\prime}]<10.5$ (top row), $9<\log_{10}[M_{\star}/h^{-1}~\mathrm{M}_{\sun}^{\prime}]<9.5$ (middle row) and $8<\log_{10}[M_{\star}/h^{-1}~\mathrm{M}_{\sun}^{\prime}]<8.5$ (bottom row).  Different colours indicate the redshift of the galaxies, as shown in the legend.  The vertical dash-dotted line in each panel indicates $\lambda_{\rm rest}=70$~$\mu$m, the approximate rest-frame wavelength shorter than which our simple dust model breaks down (hatched regions).}
\label{fig:stacked_SEDs_rest}
\end{figure}
\begin{figure*}
\centering
\includegraphics[trim = 0 0 0 0,clip = True,width = 0.5\linewidth]{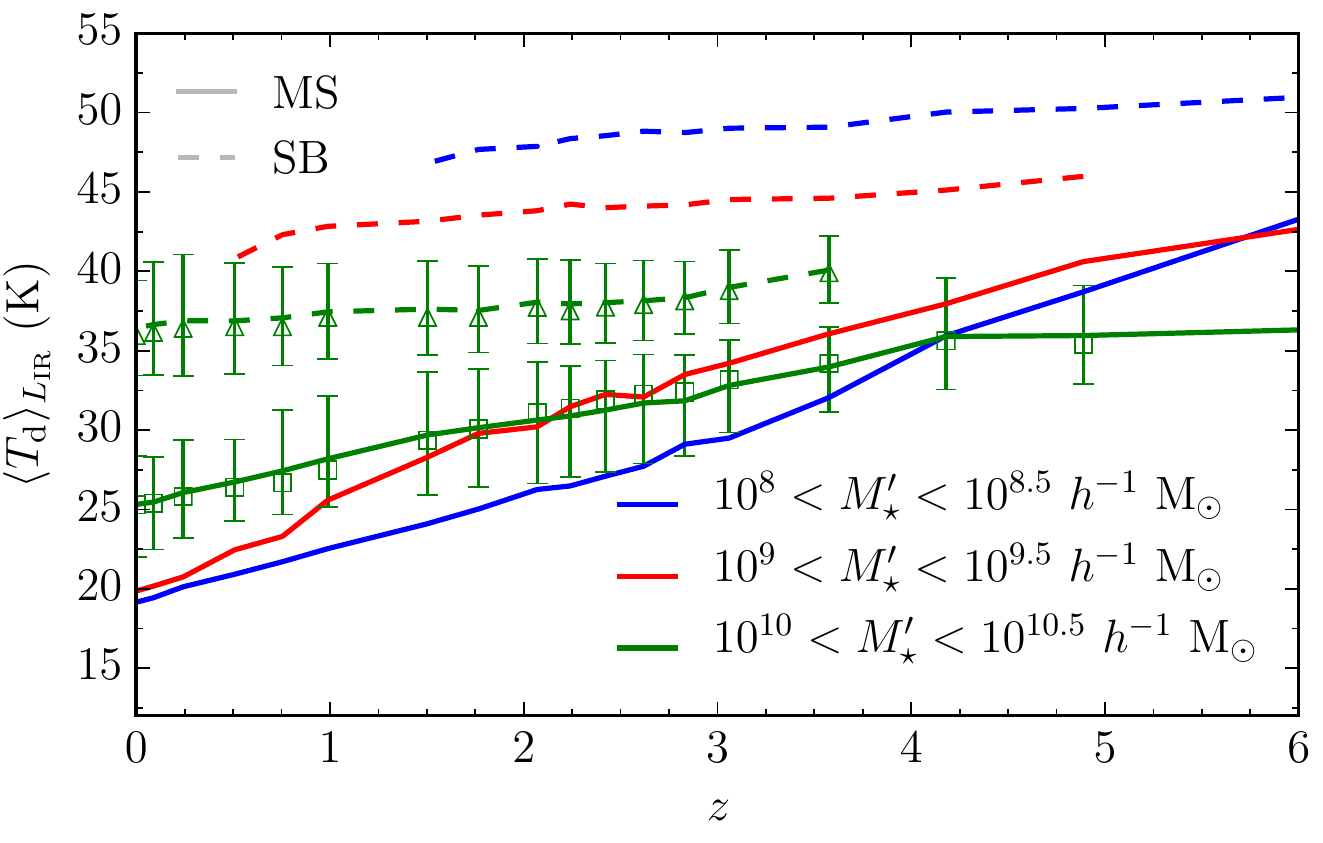}\includegraphics[trim = 0 0 0 0,clip = True,width = 0.5\linewidth]{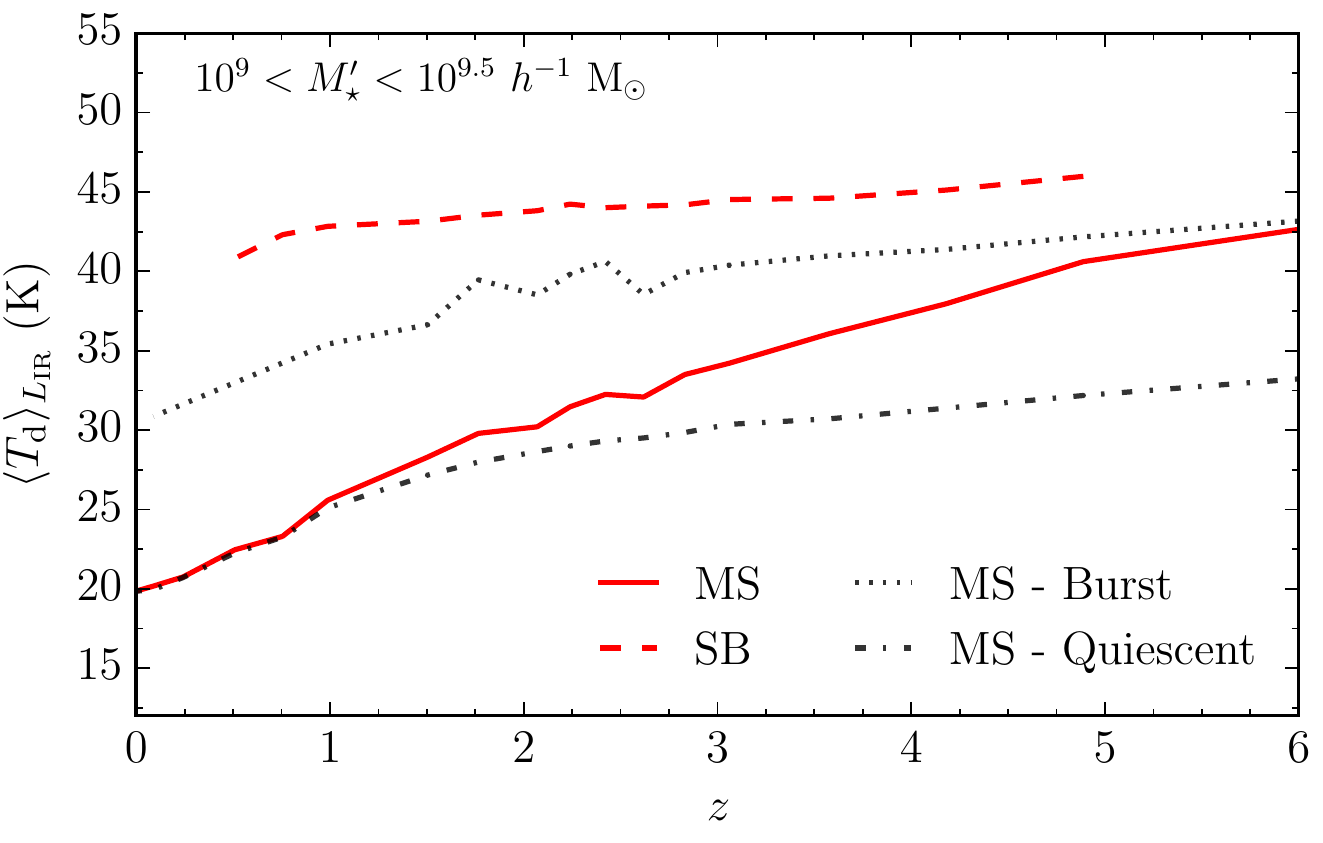}
\caption{Predicted evolution of infra-red luminosity weighted mean dust temperature.  \emph{Left panel:} For main sequence (MS, solid lines) and starburst (SB, dashed lines) galaxies.  The different colour lines indicate different stellar mass selected samples as shown in the legend. The symbols indicate the median $L_{\rm IR}$ weighted dust temperature for the highest mass sample for MS (open squares) and SB (open triangles) galaxies with the errorbars indicating the $16-84$ ($1\sigma$) $L_{\rm IR}$ weighted percentile scatter.  \emph{Right panel:} Luminosity weighted dust temperature evolution for the $10^{9}-10^{9.5}$~$h^{-1}$~M$_{\sun}$ MS sample (red solid line) and for this sample split by burst mode dominated and quiescent mode dominated galaxies (grey dotted and dash-dotted lines respectively).  The SB temperature evolution for galaxies in this mass range is also shown for reference (red dashed line).}
\label{fig:Td_z}
\end{figure*}
\begin{figure*}
\includegraphics[trim = 0 200 0 0,clip = true,width = \linewidth]{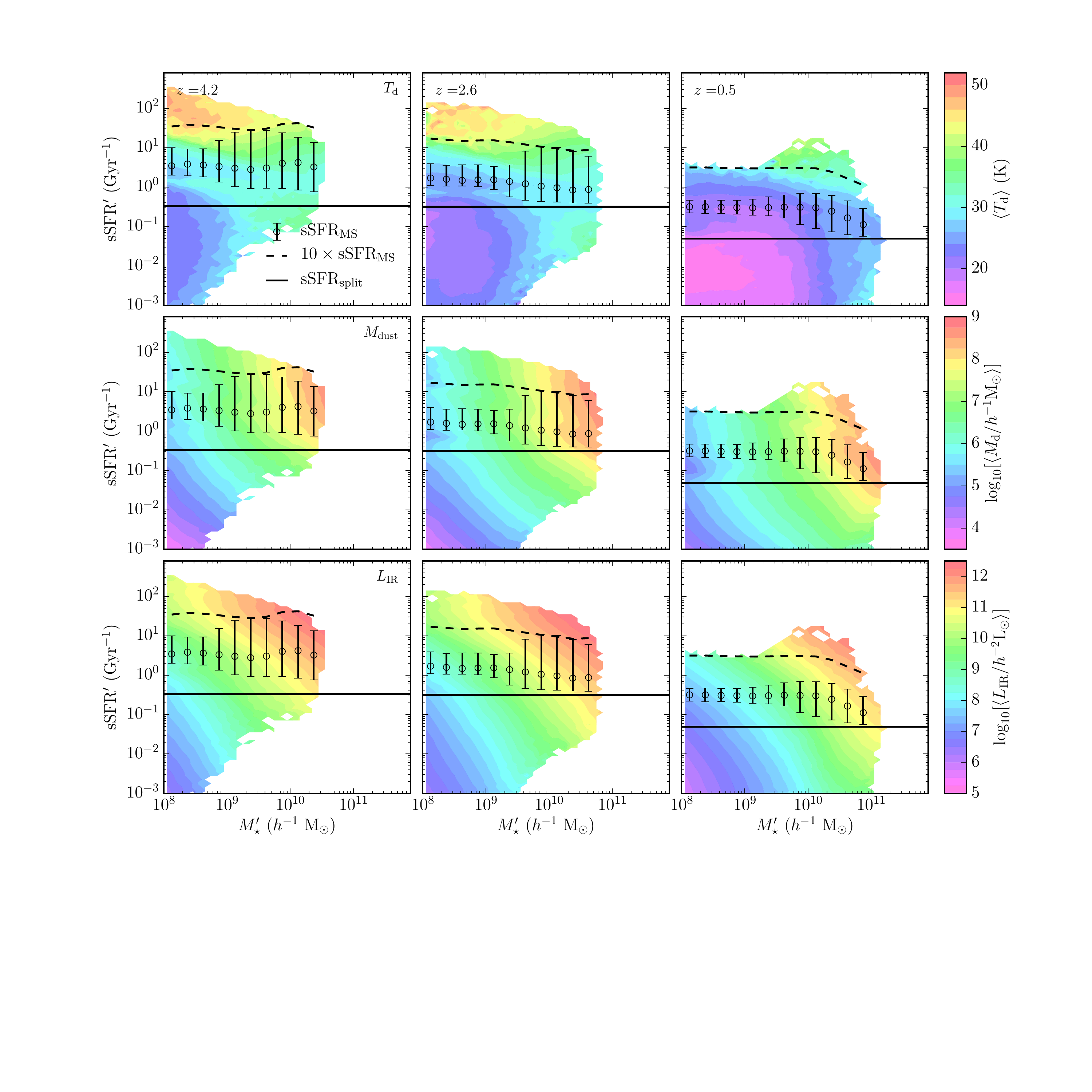}
\caption{The colour coding (see right hand side colourbar for scale) indicates the average dust temperature (top row),  dust mass (middle row) and infra-red luminosity (bottom row) at different locations in the sSFR$^{\prime}$-$M_{\star}^{\prime}$ plane, at redshifts of $4.2$ (left column), $2.6$ (middle column) and $0.5$ (right column).  Lines and symbols have the same meaning as in Fig.~\ref{fig:sSFR_SM_Planes}.}
\label{fig:Td_Md_Ld_planes}
\end{figure*}
Now we investigate the average spectral energy distributions (SEDs) at FIR wavelengths  ($8-1000$~$\muup$m) predicted by the L16 model for MS and SB galaxies as defined above.  In Fig.~\ref{fig:stacked_SEDs_rest} we show the average SEDs for both populations at a range of stellar masses and redshifts.  The broad trend of increasing bolometric luminosity with increasing redshift can be explained by the evolution of sSFR$^{\prime}_{\rm MS}$ shown in Fig.~\ref{fig:sSFR_MS_z}, that in star-forming galaxies at a fixed stellar mass, the SFRs are generally higher at higher redshift, since the bolometric infra-red luminosity $L_{\rm IR}$ closely traces the star formation rate for systems with high dust extinction.  The trend of bolometric luminosity increasing with mass, such that more massive galaxies on average have more star formation, is a simple consequence of the MS selection. The sSFR$^{\prime}_{\rm MS}$ is approximately constant over the mass ranges shown, thus higher stellar masses correspond to selecting higher star formation rates (and thus higher $L_{\rm IR}$).  For this reason, at a given redshift and stellar mass, the bolometric luminosity of the SB SEDs are higher.   

We can also see in Fig.~\ref{fig:stacked_SEDs_rest} changes in the wavelength at which the average FIR SED peaks, due to variations in the average dust temperature of the selected sample.  We show the evolution in the $L_{\rm IR}$ weighted average dust temperature of our samples in Fig.~\ref{fig:Td_z}. We weight by $L_{\rm IR}$ to reflect the temperature that will dominate the stacked SEDs.  In the top panel we see that for all samples the average dust temperature is predicted to increase with redshift.  Dust temperature is driven by the ratio of infra-red luminosity to dust mass, as $L_{\rm IR}/M_{\rm dust}\propto T_{\rm d}^{\beta+4}$ (for single temperature dust in thermal equilibrium with a dust opacity that scales as $\kappa_{\rm d}\propto\lambda^{-\beta}$).  At higher redshifts, galaxy SFRs (at a given stellar mass) are generally higher, resulting in a higher $L_{\rm IR}$, whilst the distribution of dust masses evolves much less.  This is probably due to a combination of competing effects such as the gas fractions of galaxies at a given stellar mass decreasing with redshift as cold gas is converted into stars and the metallicity of these galaxies increasing with time as stars return metals into the ISM, resulting in a dust mass (proportional to the product of cold gas mass and metallicity) that does not evolve strongly with time.  This produces the hotter dust temperatures at higher redshift. This is shown in Fig.~\ref{fig:Td_Md_Ld_planes}, where we plot the sSFR$^{\prime}$-$M_{\star}^{\prime}$ plane, but with the colourscale now indicating, from the top to bottom rows, the average dust temperature, dust mass and infra-red luminosity at that position on the plane.  We also see from the top row of Fig.~\ref{fig:Td_Md_Ld_planes} that the range of temperatures $\sim20-40$~K and temperature gradient across the main sequence (hotter dust found above the main sequence) are extremely similar to those reported by \cite{Magnelli14}.     

In the left panel of Fig.~\ref{fig:Td_z} we can see that the evolution of the average temperature is stronger for the MS samples.  This is because they are composed of both burst and quiescent mode dominated galaxies in proportions that depend on stellar mass and redshift, whereas the SB samples are predominantly populated by burst mode dominated galaxies.  In the right panel of Fig.~\ref{fig:Td_z} we illustrate this point by showing the temperature evolution for the MS (red line) and SB (red dashed line) populations for an intermediate stellar mass sample ($10^{9}-10^{9.5}$~$h^{-1}$~M$_{\sun}$) along with the evolution for the burst mode dominated and quiescent mode dominated galaxy populations in the MS sample (grey dotted and dash-dotted lines respectively).  We can see that at high redshifts the average temperature is dominated by the burst mode dominated galaxies in the MS sample, transitioning to being dominated by quiescent galaxies at low redshift.  This mixing of star formation modes on the MS could potentially be behind the sharp increase in the radiation field $\langle U\rangle$ (strongly correlated with dust temperature) found by \cite{Bethermin15} for stacked MS galaxies at $z\gtrsim2$.       

The points with errorbars in the upper panel of Fig.~\ref{fig:Td_z} show the median and the $16-84$ percentile scatter of dust temperatures for the high mass sample, ($10^{10}-10^{10.5}$~$h^{-1}$~M$_{\sun}$). The scatter for MS galaxies ($\sim10$~K) is slightly larger than for the SB population ($\sim5$~K). This reflects the broader range of SFRs in, and the contribution from both modes of star formation to, the MS sample.   

\subsection{Comparison with observations}
\label{subsec:Compare}   
\begin{figure*}
\includegraphics[trim = 0 0 0 0,clip = True,width=\linewidth]{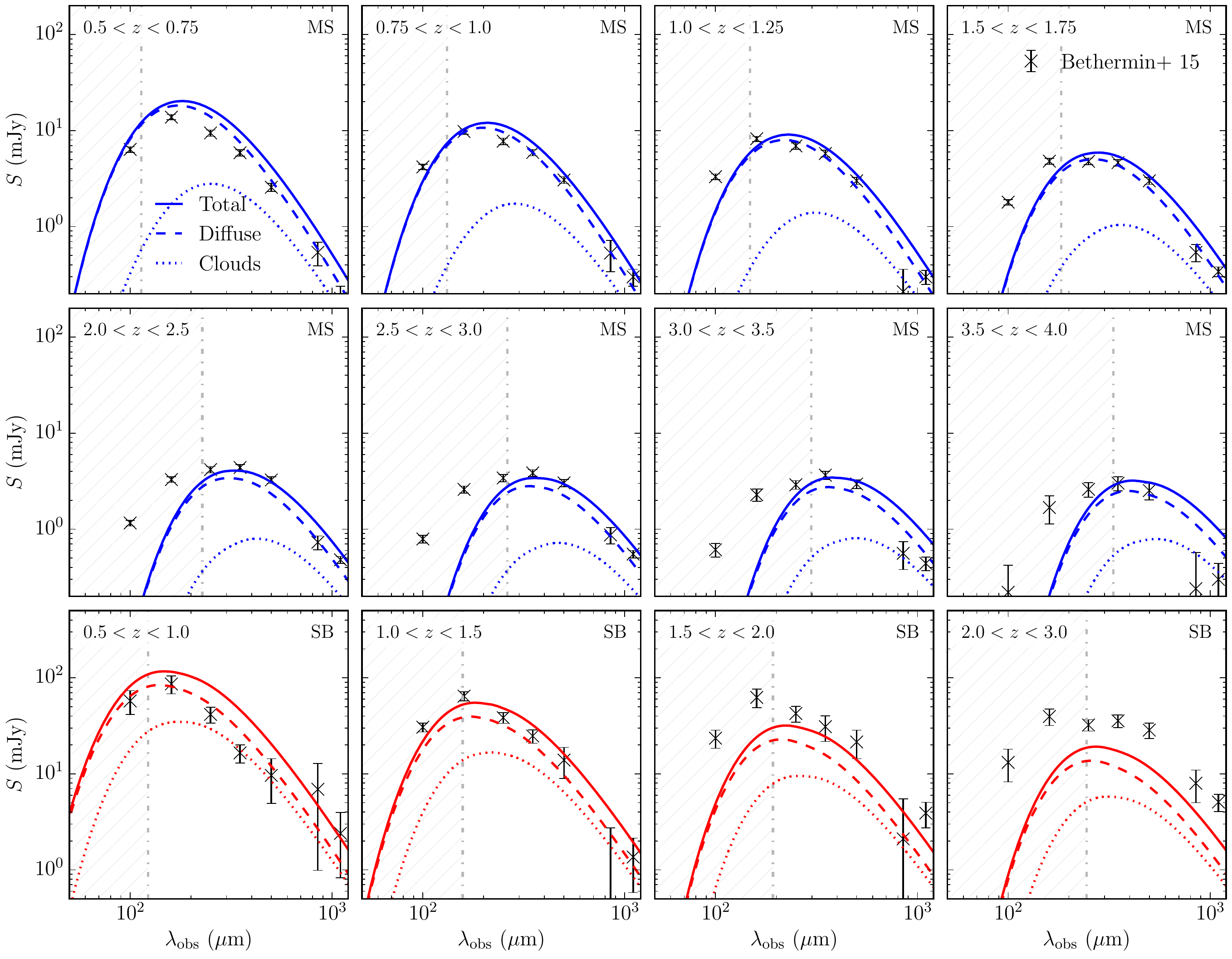}
\caption{Stacked FIR/sub-mm SEDs of main sequence galaxies (MS, top two rows) and starburst galaxies (SB, bottom row), at a range of redshift intervals indicated in each panel.  Dotted and dashed lines indicate the stacked SED for the molecular cloud and diffuse ISM dust components respectively.  Observational data (crosses with errorbars) are from Bethermin et al. (\citeyear{Bethermin15}). The vertical dash-dotted line in each panel indicates $\lambda_{\rm rest}=70$~$\mu$m, the approximate wavelength shorter than which our simple dust model breaks down (hatched regions).  For presentation purposes, a representative subset of the Bethermin et al. redshifts intervals are displayed here.}
\label{fig:Bethermin_SEDs}
\end{figure*}
\begin{figure}
\includegraphics[trim = 0 0 0 0,clip = True,width=\linewidth]{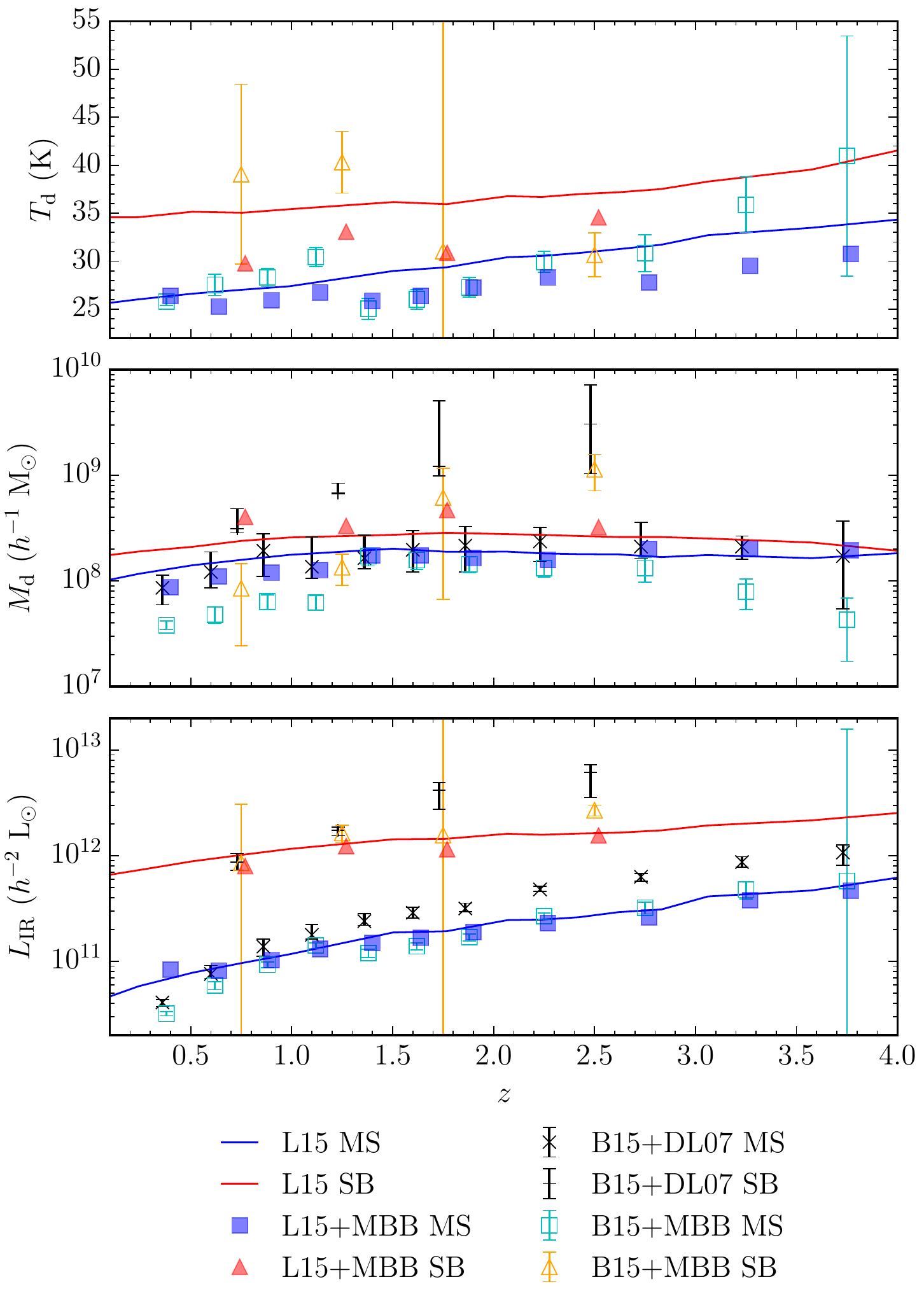}
\caption{Physical properties, dust temperature ($T_{\rm d}$, top panel), dust mass ($M_{\rm d}$, middle panel) and total infra-red dust luminosity ($L_{\rm IR}$, bottom panel) derived from a modified blackbody (equation~\ref{eq:MBB}) fit to both the simulated stacked photometry from the L16 model (filled symbols) and the observed stacked photometry from Bethermin et al. (\citeyear{Bethermin15}, B15, open symbols), for main sequence (MS, blue/cyan) and starburst (SB, red/orange, model/observations) populations respectively.  Also shown for reference are the average values for each population predicted directly by the L16 model (solid lines), and the values predicted by fitting the dust model of Draine \& Li (\citeyear{DL07}) to the observed photometry, as was done in B{\'e}thermin et al. for MS (crosses) and SB (bars) populations.}
\label{fig:MBB_fit}
\end{figure}
We now perform a detailed comparison of the predictions of the model with the observational results of \cite{Bethermin15}.  These authors stacked infra-red images of a stellar mass selected sample of galaxies taken from the \cosmos field, using a stellar mass limit of $3\times10^{10}$~M$_{\sun}$ so that their sample was complete up to $z\sim4$. They also removed X-ray detected active galactic nuclei (AGN) hosts, so we do not consider an AGN component in our simulated SEDs.  We use the same mass limit as Bethermin et al. [scaled to a universal \cite{Kennicutt83} IMF as described earlier] and consider galaxies in the same redshift bins.  In our simulation we stack galaxy SEDs over a redshift interval by weighting averaged SEDs at each output redshift by the comoving number density of selected galaxies, $n(z)$, and the comoving volume element $\mathrm{d}V/\mathrm{d}z$. The average SED for each redshift interval is therefore calculated using 
\begin{equation}
S_{\lambda_{\rm o}}=\ddfrac{\int_{z_{1}}^{z_{2}}\frac{(1+z)\langle L_{\lambda_{\mathrm{o}}/(1+z)}\rangle(z)}{4\pi D_{\rm L}^{2}(z)}n(z)\frac{\mathrm{d}V}{\mathrm{d}z}\mathrm{d}z}{\int_{z_{1}}^{z_{2}}n(z)\frac{\mathrm{d}V}{\mathrm{d}z}\mathrm{d}z}\mathrm{.}
\end{equation}  
Here, $S_{\lambda_{\rm o}}$ is the flux at some observer-frame wavelength $\lambda_{\rm o}$ which is related to the corresponding emitted (rest-frame) wavelength $\lambda_{\rm e}$ by $\lambda_{\rm e}=\lambda_{\rm o}/(1+z)$, $\langle L_{\lambda_{\rm o}/(1+z)}\rangle$ is the average luminosity of the sample at this rest-frame wavelength, $D_{\rm L}$ is the luminosity distance to redshift $z$, and $z_{1}$ and $z_{2}$ represent the lower and upper limit respectively of the redshift bin considered.

We show the comparison to the data in Fig.~\ref{fig:Bethermin_SEDs}.  The agreement between the model predictions and the observations for MS galaxies is extremely good (for $z\gtrsim0.5$), which is remarkable given that the SEDs of galaxies were not considered in calibrating the model. For SB galaxies (bottom row) the agreement is generally good for $z\lesssim2$.   

To further investigate how our predictions compare to the observations, we compare their dust temperatures, dust masses and infra-red luminosities as a function of redshift in Fig.~\ref{fig:MBB_fit}.  To do this in a consistent way, we fit a modified black body (MBB),
\begin{equation}
L_{\lambda} = 4\pi\,M_{\rm dust}\,\kappa_{0}(\lambda/\lambda_{0})^{-\beta}\,B_{\lambda}(T_{\rm d})\rm{,}
\label{eq:MBB}
\end{equation}
to both the observed and predicted photometry, noting that this form is not equivalent to the one assumed for the dust emission in \galform (Equation~\ref{eq:galform_dust_emission}) and has only a single dust temperature (whereas the \galform dust model assumes that there are two temperatures).  We choose to assume a fixed $\beta=2$ and only consider wavelengths from the available photometry ($100$, $160$, $250$, $350$, $500$, $850$ and $1100$~$\muup$m) for which $\lambda_{\rm rest}>70$~$\muup$m, so that we are confident that the approximations in our dust model (and in using the MBB) are valid.  In order to derive a dust mass from the MBB fits we must assume an opacity for the dust. Here we use $\kappa_{0}=6.04$~cm$^{-2}$~g$^{-1}$ at $\lambda_{0}=250$~$\muup$m, such that the opacity per unit mass of metals in the gaseous phase of \cite{DraineLee84} is regained for our value of $\delta_{\rm dust}=0.334$.  

For the observed photometry, we calculate the errors on the physical properties using the method of \cite{Magdis12}. Using the original flux measurements and measurement errors we generate $1000$ simulated flux sets using a Gaussian distribution, and fit an MBB to these in the same way.  The standard deviations in the derived values are then taken to represent the uncertainty in the values derived from the original observed photometry.
 
For MS galaxies the agreement in $\Td$ is generally good up to $z\sim3$, at higher redshifts the observations appear to favour higher dust temperatures.  This is a consequence of the higher dust masses predicted by the model, as the infra-red luminosities are in good agreement.  This suggests that the model overproduces dust at these high redshifts. This is due, at least in part, to the top heavy IMF in burst star formation and the abundance of burst mode MS galaxies at these higher redshifts.  When we repeat this calculation with the GP14 model, which has a universal IMF, we find MS dust masses at $3.5\lesssim z\lesssim4$ that are a factor of $\sim2$ lower.  However, this model does not reproduce the observations as well over the whole redshift range considered.  

It is also possible that systems at higher redshifts are composed of more heterogeneous dust distributions than are accounted for in both \galform and the MBB fit, reflected in the larger errorbars for the fit to the observed photometry at higher redshifts, meaning that these physical properties are poorly constrained.  However, the larger errors could also be due to our restriction of having $\lambda_{\rm rest}>70$~$\muup$m for the MBB fit, which means that our highest redshift bin has only $4$ data points in the SED.     

For SB galaxies, the model and the data are in good agreement for $z\lesssim2$. At higher redshifts, the model's average infra-red luminosity appears too low to reproduce the observed photometry, as also seen in Fig~\ref{fig:Bethermin_SEDs}.  Given the mass selected nature of the observed sample it is unlikely that this is caused by gravitational lensing, which is not included in the model, boosting the observed flux.  One could imagine the dust geometry/composition of these extreme SB galaxies being more complicated than is modelled in \galform, however in the model these galaxies are in the regime where $\gtrsim95$~per~cent of the stellar luminosity is being re-radiated by dust\footnote{For a sample of SB galaxies at $z\sim2$, for MS galaxies the percentage is $\sim65-95$.}, so it is unlikely that assuming more complicated geometry could account for the difference in $L_{\rm IR}$ seen between the model and observations.  It is therefore more likely to be a result of the inferred SFRs in the model being too low.  The number of observed galaxies in the highest redshift SB bin is also relatively small $\sim5$, so given that the observations could also be affected by sample variance, the significance of the discrepancy between the model predictions and the observations here is unclear.  Performing a similar observational study over larger areas than the $\sim1$~deg$^{2}$ used by \cite{Bethermin15} (and/or using multiple fields) would help shed light onto the significance of these extremely IR luminous objects.   

Of note also in Fig. \ref{fig:MBB_fit} are instances where the value derived from the MBB fit to the simulated stacked photometry fails to reproduce the average value produced directly by the model (solid lines).  We attribute these to the fact that the MBB makes different assumptions about the dust emission than are made in the model.  The most striking example of this is in the dust temperatures of SB galaxies.  SB galaxies are essentially all undergoing burst mode star formation, in the model this means a value of $\beta_{\rm b}=1.5$ is used, for rest-frame wavelengths longer than $\lambda_{\rm b}=100$~$\muup$m (see equation \ref{eq:kappa_d}), to predict the dust SED. However, we assume a value of $\beta = 2$ in the MBB fit (see equation \ref{eq:MBB}).  This causes the MBB to return cooler dust temperatures.  

There are also instances (predominantly at $z\lesssim1.5$) where the MBB fit fails to reproduce the average dust mass predicted by the model.  This we attribute to having two dust components in our model, whereas the MBB, by construction, assumes only one.  Our model SEDs are dominated by the diffuse dust component, as this has a much greater luminosity than the molecular clouds (Fig.~\ref{fig:Bethermin_SEDs}).  However, as $f_{\rm cloud}=0.5$, in our model both components have the same dust mass.  The MBB fit therefore `misses' much of the less luminous dust in clouds.  If we change the parameters of our dust model such that $\beta_{\rm b}=2$ and $f_{\rm cloud}=0$ then these discrepancies disappear.  This highlights a drawback of using MBB fits to derive physical properties from dust SEDs.  Even in the simplified case where a galaxy's dust SED is the sum of two modified blackbodies (as is the case in \galform) and the `correct' dust opacity is used in the MBB fit, the MBB blackbody does not always return the correct values for the physical properties relating to the dust SED.  In the worst case here the MBB underpredicts the actual average temperature by $\sim5$~K.

For reference we have also shown in Fig.~\ref{fig:MBB_fit} the values derived from fitting the dust model of \cite{DL07} to the observed photometry as is done in \cite{Bethermin15}.  The MBB fits tend to produce slightly lower $L_{\rm IR}$ values, typically by small factors $\sim1-2$, compared to the Draine \& Li model values.  This is easily understood by the Draine \& Li model allowing for emission in the mid- and near- infrared.  We can also see that the MBB tends to predict lower dust masses than the Drain \& Li model, as found by \cite{Magdis12}.  These differences highlight the caution that is required when interpreting physical properties derived from modelling dust SEDs.

The results in this section suggest that the L16 model can accurately predict the average dust emission of MS galaxies, which contribute to the bulk of the SFR density, over a broad range of redshifts.
\subsection{Constraints on dust model parameters}
\label{subsec:dust_params}
\begin{figure*}
\includegraphics[trim = 0 16.25 0 0,clip = True,width = 0.5\linewidth]{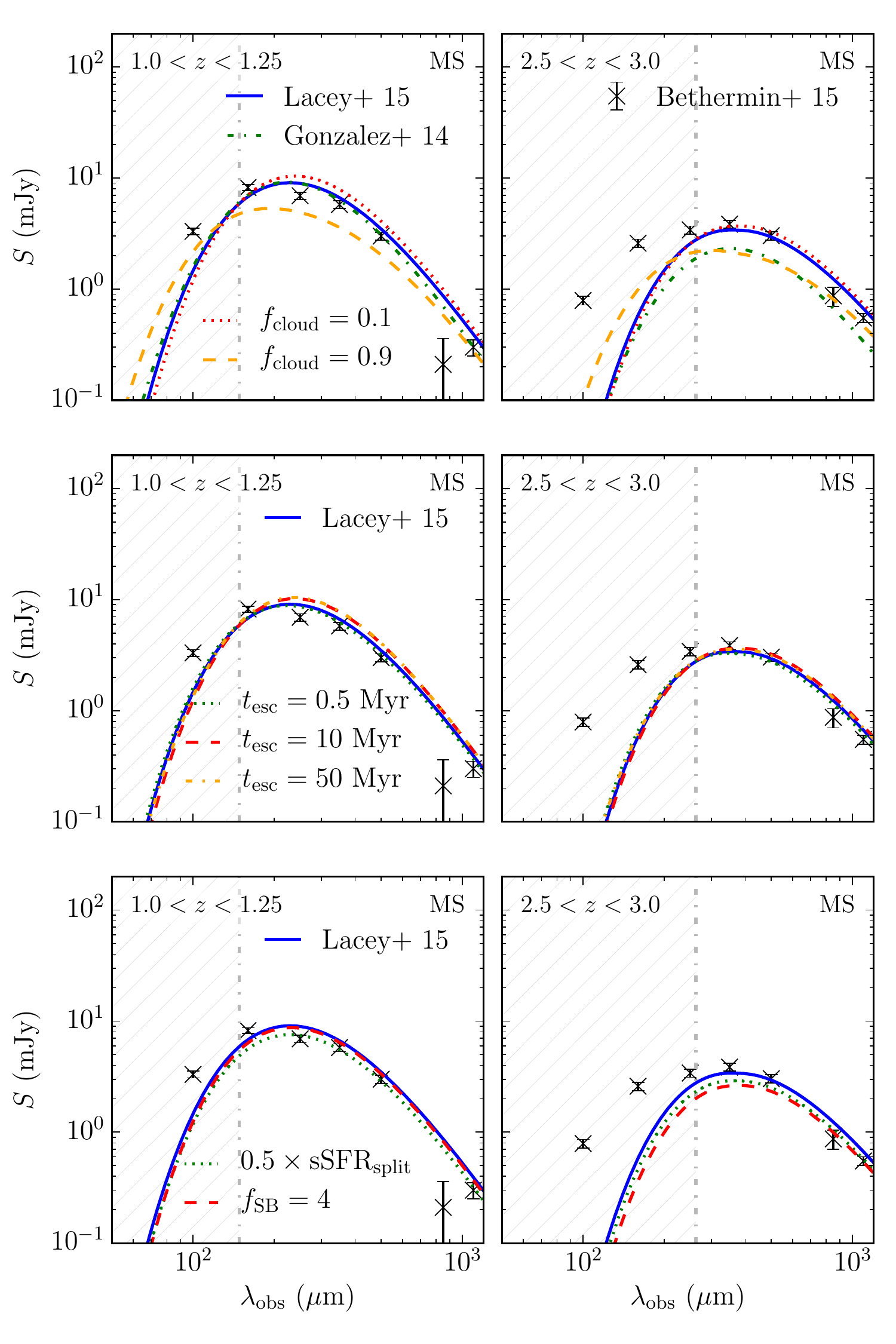}\includegraphics[trim = 0 16.25 0 0,clip = True,width = 0.5\linewidth]{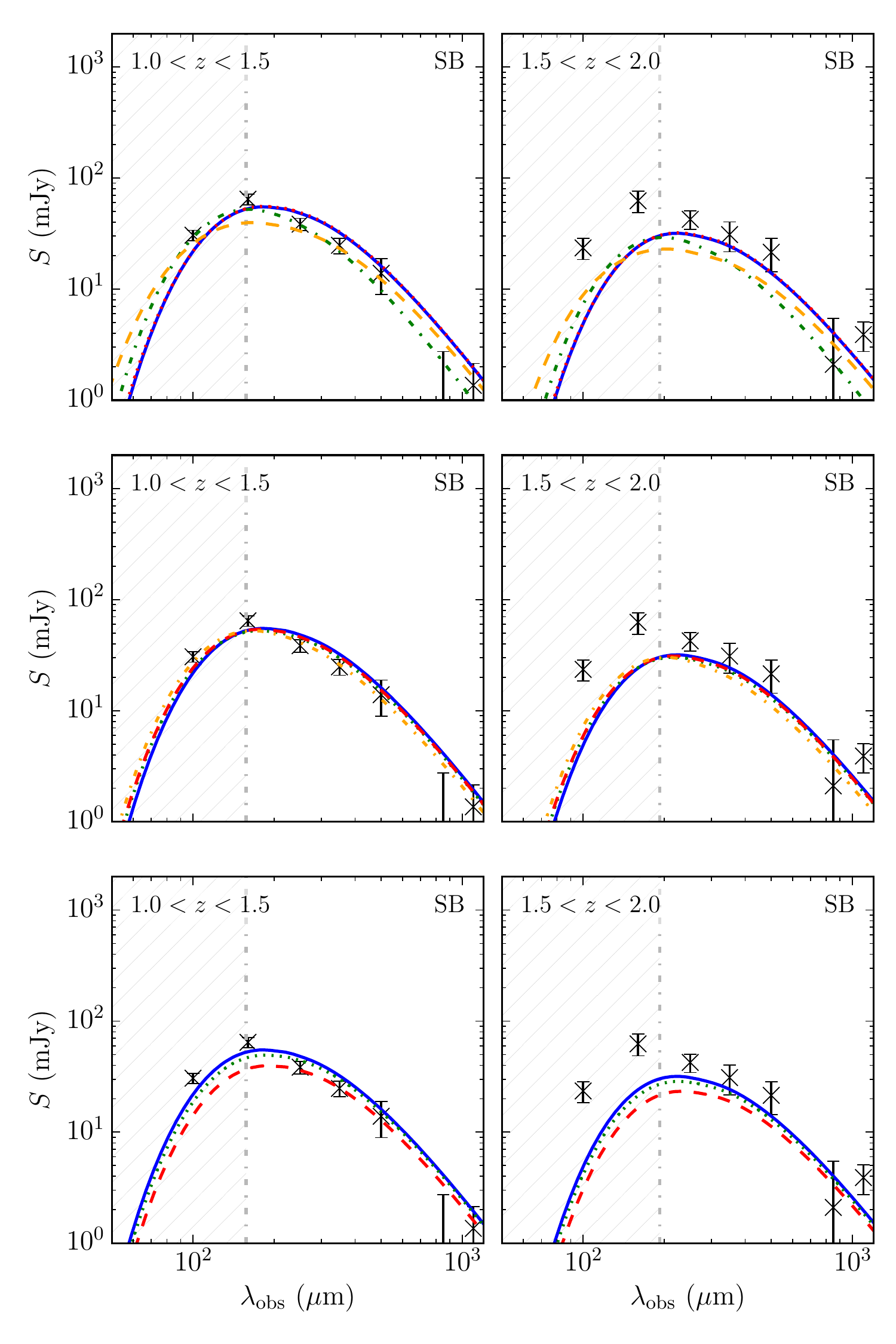}
\caption{The effect of varying model parameters and assumptions on the predicted stacked SEDs for MS (left two columns) and SB (right two columns) galaxies for the redshift intervals indicated in the panels.  \emph{Top row:} Predictions for the fiducial L16 model (blue solid line), the GP14 model (green dash-dotted line) and the effect of varying the dust model parameter $f_{\rm cloud}$ on the L16 model predictions (the fiducial value of this parameter is $0.5$).  \emph{Middle row:} Effect of varying $t_{\rm esc}$ on the L16 model predictions (the fiducial value is $1$~Myr).  \emph{Bottom row:} Effect of varying the value of sSFR$_{\rm split}$ and $f_{\rm SB}$.  The vertical dashed-dotted line in each panel indicates a rest-frame wavelength of $70$~$\muup$m, shorter than which our simple dust model breaks down (hatched regions).  The observational data are from B\'{e}thermin et al. (\citeyear{Bethermin15}, crosses with errorbars).}
\label{fig:Bethermin_SEDs_Compare}
\end{figure*}
\begin{figure}
\includegraphics[width=\linewidth]{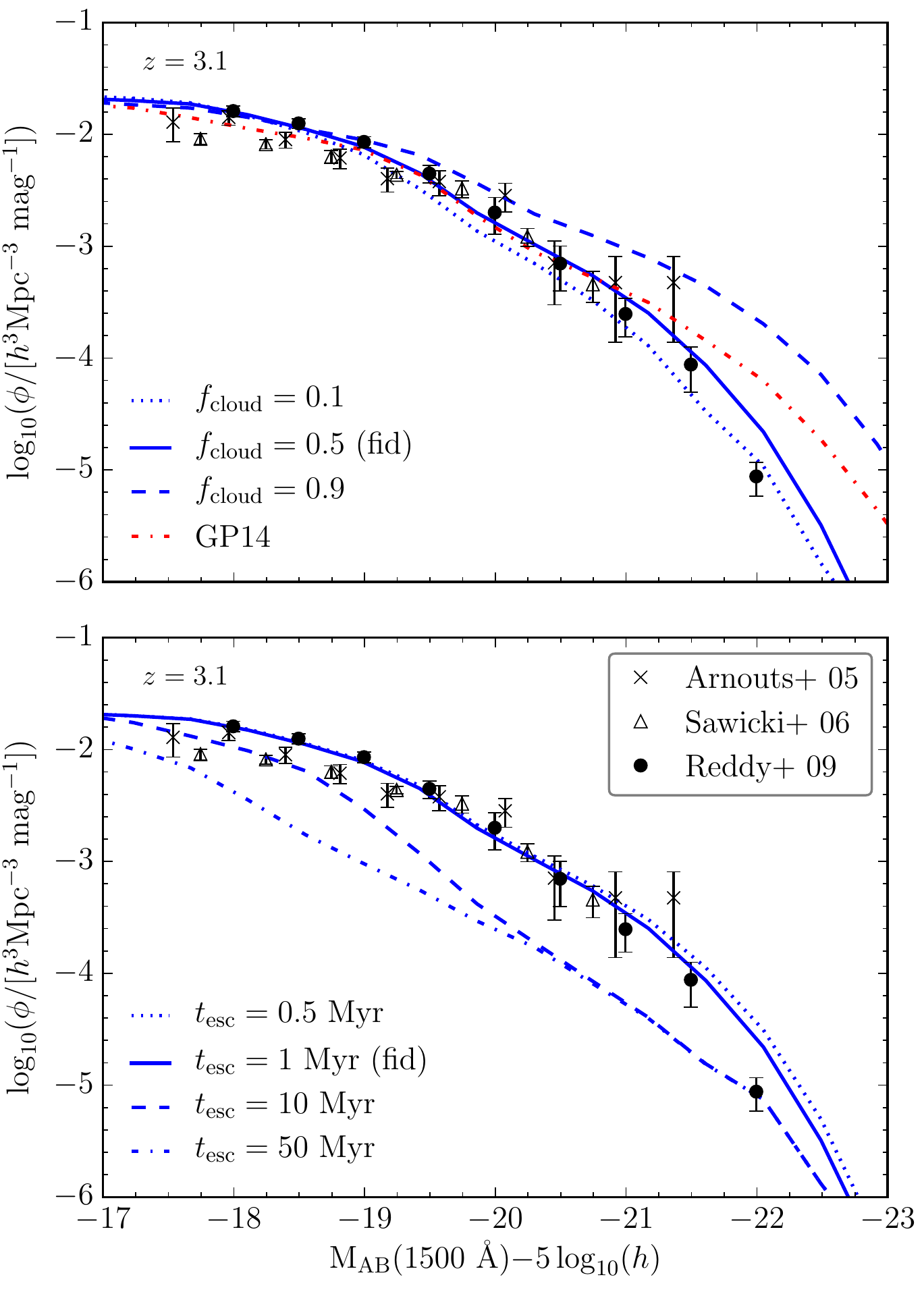}
\caption{Effect of varying parameters on the predicted rest-frame 1500~{\AA} luminosity functions at $z=3.1$.  \emph{Top panel:} We show the luminosity function for the fiducial model ($f_{\rm cloud}=0.5$, blue solid line), for $f_{\rm cloud}=0.1$ (blue dotted line) and for $0.9$ (blue dashed line) as well as for the Gonzalez-Perez et al. (\citeyear{vgp14}) model (red dash-dotted line).  \emph{Bottom panel:} We show the luminosity function for the fiducial model ($t_{\rm esc}=1$~Myr, blue solid line) and for $t_{\rm esc}=0.5$~Myr (dotted line), $t_{\rm esc}=10$~Myr (dashed line) and $t_{\rm esc}=50$~Myr (dash-dotted line).  Observational data shown in both panels are from Arnouts et al. (\citeyear{Arnouts05}, crosses), Sawicki et al. (\citeyear{Sawicki06}, open triangles) and Reddy et al. (\citeyear{Reddy09}, filled circles). }
\label{fig:LBGlfs}
\end{figure}   
In this section we consider varying parameters in the dust model to investigate the robustness of our predictions, and whether the observed average SEDs can constrain the values of these parameters.  

The parameters that we choose to vary are $f_{\rm cloud}$, the fraction of dust mass in molecular clouds, and $t_{\rm esc}$, the timescale over which stars migrate out of their birth clouds.  The fiducial values for these parameters are $f_{\rm cloud}=0.5$ and $t_{\rm esc}=1$~Myr. Here we consider values of $0.1$ and $0.9$ for $f_{\rm cloud}$ and $0.5$, $10$ and $50$~Myr  for $t_{\rm esc}$.  We choose these values as we believe they describe a physically acceptable range for these parameters \citep[see e.g. Table 2 in][]{Silva98} with the current fiducial values in the model being primarily constrained by the far-UV luminosity function of Lyman break galaxies (see below).

We choose not to explore variations in the gas mass ($m_{\rm cloud}=10^{6}$~M$_{\sun}$) and radius ($r_{\rm cloud}=16$~pc) of the clouds.  These only enter into the calculation as the ratio $m_{\rm cloud}/r_{\rm cloud}^{2}$ which, along with the gas metallicity, determines the optical depth of the clouds.  This is large at UV wavelengths, and so variations of $m_{\rm cloud}$ and $r_{\rm cloud}$ have a minor effect on the results of our simple dust model, provided the clouds are still in the optically thick regime for UV/optical light\footnote{We remind the reader that we have assumed that dust is optically thin to its own emission at FIR/sub-mm wavelengths.}.  As shown in \cite{Vega05} the main effect of changing these parameters is seen in the mid-IR dust emission, which we do not consider in this study.     

We derive SEDs for the $f_{\rm cloud}$ and $t_{\rm esc}$ variants for comparison with the \cite{Bethermin15} data as described above.  These are shown for a selection of redshifts in the top two rows of Fig.~\ref{fig:Bethermin_SEDs_Compare}, for MS galaxies (left panels), and for SB galaxies (right panels).  We also compare the observations to predictions from the GP14 model (top row), which is similar to the L16 model used predominantly throughout this work but does not have a top-heavy IMF in burst mode star formation. We also test the sensitivity of our results to parameters used in defining our galaxy populations, sSFR$_{\rm split}$, the specific star formation rate that separates passive and star-forming galaxies, and $f_{\rm SB}$, the factor above the sSFR$_{\rm MS}$ which separates MS and SB galaxies (Fig.~\ref{fig:Bethermin_SEDs_Compare}, bottom row).    

Varying the dust parameters $f_{\rm cloud}$ and $t_{\rm esc}$ results in only fairly modest changes to the predicted SEDs, for the stellar mass limit used here ($1.7\times10^{10}$~$h^{-1}$~M$_{\sun}$).  Changing the parameter $f_{\rm cloud}$ does not change the energy absorbed by the cloud component as all clouds are assumed to have a fixed mass and radius, and thus a fixed  optical depth as $\tau_{\rm cloud}\propto Z_{\rm cloud}m_{\rm cloud}/r_{\rm cloud}^{2}$, where $Z_{\rm cloud}$ is the gas metallicity of the molecular cloud.  However, it does affect the cloud dust temperature as it changes the total mass of dust in the cloud component. Hence, increasing $f_{\rm cloud}$ will make the cloud dust temperature cooler.  It also changes the mass of dust in, and thus the optical depth of, the diffuse component.  Increasing $f_{\rm cloud}$ reduces the amount of diffuse dust, thus lowering its optical depth, and therefore the amount of energy it absorbs.  How this changes the dust temperature of the diffuse component depends on the whether the reduction of energy absorbed (reducing the temperature) or reduction of dust mass (increasing the temperature) is the dominant effect.  

Increasing the escape time of stars from their birth clouds allows more energy from stellar radiation to be deposited in the cloud component of our model, increasing the amount of energy absorbed by this component and thus its dust temperature.  The diffuse component becomes cooler as less energy is then left to be absorbed by the same mass of diffuse dust.  

However, as varying these parameters has such a modest impact on the model predictions, though it should be noted that high values of $f_{\rm cloud}$ appear unlikely for main sequence galaxies, we conclude that these observations do not provide a stronger constraint on the parameters in our dust model than previously available data, such as the $1500$~{\AA} luminosity function (see Fig.~\ref{fig:LBGlfs}).  The rest-frame far-UV ($1500$~\AA) luminosity function probes star-forming galaxies typically selected at high redshifts by the Lyman-break technique, providing a strong constraint on the dust model at high redshift.  Increasing the obscuration of young stars through either increasing $t_{\rm esc}$ such that stars spend longer in their birth cloud, or decreasing $f_{\rm cloud}$ which increases the amount of energy absorbed by the diffuse component (and thus the total amount of stellar radiation absorbed), reduces the number density of objects at the bright end of the luminosity function (see Lacey et al. \citeyear{Lacey11} and Gonzalez-Perez et al. \citeyear{vgp13} for detailed studies of the effects of dust obscuration on UV selected galaxies in \galform models).   

The GP14 model makes similar predictions for the stacked SEDs to the L16 model, though it appears that it has generally lower $L_{\rm IR}$ for $z\gtrsim1$ MS galaxies, a result of the lower sSFR$_{\rm MS}$ predicted by the GP14 model for high mass galaxies (see Fig.~\ref{fig:sSFR_MS_z}).  The SEDs predicted by the GP14 model also appear to have a slightly steeper Rayleigh-Jeans tail, which is due to the choice of a larger value for $\beta_{\rm b}=1.6$ in that model, compared with the value of $\beta_{\rm b}=1.5$ used in the L16 model (see Equation~\ref{eq:kappa_d}).

Finally, in the bottom row of Fig.~\ref{fig:Bethermin_SEDs_Compare} we show predictions for the stacked SEDs predicted by the L16 model but where the value of sSFR$_{\rm split}$ was halved prior to defining the position of the main sequence on the sSFR-M$_{\star}$ plane, and where the factor $f_{\rm SB}$, which controls the divide between SB and MS galaxies was reduced from its fiducial value of $10$ to $4$.  We note that neither of these changes makes a significant difference (which is reassuring as it means our results are not sensitive to choices we have made in defining the MS and SB populations) other than to slightly lower the normalisation of the SEDs and shift the peak to slightly longer wavelengths. This is because these changes will generally result in slightly lower SFRs and cooler dust temperatures (see Fig.~\ref{fig:Td_Md_Ld_planes}) being selected in the MS and SB populations. 

\section{Conclusions}
\label{sec:conclusion}
The re-emission of radiation by interstellar dust produces a large proportion of the extragalactic background light, implying that a significant fraction of the star formation over the history of the Universe has been obscured by dust.  Understanding the nature of dust absorption and emission is therefore critical to understanding galaxy formation and evolution.

However, the poor angular resolution of most current telescopes at the FIR/sub-mm wavelengths at which dust emits ($\sim$~$20$~arcsec FWHM) means that in the FIR/sub-mm imaging only the brightest galaxies (with the highest SFRs) can be resolved as point sources above the confusion background. These galaxies comprise either starburst galaxies which lie above the main sequence of star-forming galaxies on the sSFR-$M_{\star}$ plane, and do not make the dominant contribution to the global star formation budget, or the massive end (e.g. $M_{\star}\gtrsim10^{10.5}$~$h^{-1}$~M$_{\sun}$ at $z\approx2$) of the main sequence galaxy population.  For less massive galaxies, and at higher redshifts, where the galaxies cannot be resolved individually in the FIR/sub-mm imaging, their dust properties can be investigated through a stacking analysis, the outcome of which is an average FIR/sub-mm SED.      

We present predictions for such a stacking analysis from a state-of-the-art semi-analytic model of hierarchical galaxy formation.  This is coupled with a simple model for the reprocessing of stellar radiation by dust in which the dust temperatures for molecular cloud and diffuse dust components are calculated based on the equations of radiative transfer and energy balance arguments, assuming the dust emits as a modified blackbody.  This is implemented within a $\Lambda$CDM Millennium style $N$-body simulation which uses the \emph{WMAP7} cosmology.

In a way consistent with observations, we define two populations of star-forming galaxies based on their location on the sSFR$^{\prime}$-$M_{\star}^{\prime}$ plane [where the prime symbol represents the value for this physical property that would be inferred assuming a universal Kennicutt \citeyear{Kennicutt83} IMF, see Section~\ref{subsubsec:infer}], namely main sequence (MS) if they lie close to the main locus of star-forming galaxies and starburst (SB) if they are elevated on that plane relative to the MS.  We note that these definitions do not necessarily reflect the quiescent and burst modes of star formation as defined within the model based on physical criteria.  Quiescent mode star formation takes place within the galaxy disc, and follows an empirical relation in which the star formation depends on the surface density of molecular gas in the disc.  Burst mode star formation takes place in the bulge after gas is transferred to this from the disc by some dynamical process, either a merger or a disc instability.  Burst mode dominated galaxies have generally hotter dust temperatures (driven by their enhanced SFRs) than quiescent mode dominated galaxies.  Our model incorporates a top-heavy IMF, characterised by a slope of $x=1$, for star formation in burst mode.  However, when we make comparisons to physical properties we scale all quantities to what would be inferred assuming a universal \cite{Kennicutt83} IMF (see Section~\ref{subsubsec:infer}).  Most conversion factors are taken from the literature and are described in the text.  However, we do not apply a conversion factor to the true stellar masses predicted by our model, despite the assumption of a top-heavy IMF for burst mode star formation.  As discussed in Appendix~\ref{app:IMF} this has a relatively small effect on the stellar masses that would be inferred fitting the UV/optical/near-IR SED, a technique commonly used in observational studies, compared to the uncertainties and/or scatter associated with this technique.

The model exhibits a tight main sequence (sSFR$^{\prime}=$sSFR$^{\prime}_{\rm MS}$) on the sSFR$^{\prime}$-$M^{\prime}_{\star}$ plane when galaxies are able to self-regulate their SFR through the interplay of the prescriptions for gas cooling, quiescent mode star formation and supernovae feedback.  In instances where this is not the case through either (i) dynamical processes triggering burst mode star formation, (ii) environmental processes such as ram-pressure stripping limiting gas supply or (iii) energy input from AGN inhibiting gas cooling, this causes the scatter around sSFR$^{\prime}_{\rm MS}$ to increase.  We observe a negative high mass slope for sSFR$^{\prime}_{\rm MS}$ at low redshifts ($z\lesssim1$) which we attribute to AGN feedback in high mass halos.  This is also reflected in high bulge-to-total mass ratios in these galaxies.  This negative slope exists at higher redshifts in quiescent mode dominated galaxies but is not seen for the total galaxy population, because at these redshifts the high mass end of the main sequence is populated predominantly by burst mode dominated galaxies.  Additionally we find the model predicts that galaxies classified as being on the main sequence make the dominant contribution to the star formation rate density at all redshifts, as is seen in observations. For redshifts $z\gtrsim2$ this contribution is predicted to be dominated by galaxies that lie on the main sequence but for which the current SFR is dominated by burst mode star formation.   

We investigate the redshift evolution of the average temperature for main sequence galaxies and find that it is driven primarily by the transition from the main sequence being dominated by burst mode star formation (higher dust temperatures) at high redshifts, to quiescent mode star formation (lower dust temperatures) at low redshifts.

We compare the average (stacked) FIR SEDs for galaxies with $M^{\prime}_{\star}>1.7\times10^{10}$~$h^{-1}$~M$_{\sun}$ at a range of redshifts with observations from \cite{Bethermin15}.  For main sequence galaxies the agreement is very good for $0.5<z<4$. The model predicts dust temperatures in agreement with those inferred from observations accurately up to $z\sim3$, while at higher redshifts the observations appear to favour hotter dust temperatures than the model predicts.  This appears to be due primarily to the model producing too much dust at these redshifts.  It could also be that real galaxies are more heterogeneous at higher redshifts e.g. clumpier dust distributions resulting in a range of dust temperatures, which would not be well captured by our simple dust model.  

For starburst galaxies, which lie elevated relative to the main sequence on the sSFR$^{\prime}$-$M_{\star}$ plane, the agreement between the model and observations is also encouraging for $0.5\lesssim z\lesssim2$.  For $z\gtrsim2$ the model appears to underpredict the average $L_{\rm IR}$ inferred from the observations.  This implies that the model does not allow enough star formation at higher redshifts ($z\gtrsim2$) in extremely star-forming systems.  However, the model \emph{is} calibrated to reproduce the observed $850$~$\muup$m number counts, which are composed predominantly of galaxies at $z\sim1-3$ undergoing burst mode star formation.  The apparent discrepancy here is most probably due to how these populations are defined.  As we have shown, many of the model galaxies undergoing burst mode star formation at $z\gtrsim2$ would be classified as MS based on their position on the sSFR$^{\prime}$-$M_{\star}^{\prime}$ plane, and their SEDs not included in the SB stack.  Thus the model can underpredict the average SEDs of objects with extreme sSFRs at high redshifts whilst still reproducing the abundance of galaxies selected by their emission at $850$~$\muup$m at similar redshifts.             

We investigate whether the predictions for the stacked SEDs are sensitive to choices made for the values of parameters in our dust model, mainly the fraction of dust in molecular clouds ($f_{\rm cloud}$) and the escape time of stars from their molecular birth clouds ($t_{\rm esc}$).  We find that varying these parameters causes only fairly modest changes to the predicted stacked SED, thus these observational data do not provide a stronger constraint on these parameters than previously available data, e.g. the rest-frame $1500$~{\AA} luminosity function at $z\sim3$.  

In summary, the predictions made by our simple dust model, combined with our semi-analytic model of galaxy formation provide an explanation for the evolution of dust temperatures on the star-forming galaxy main sequence, and can reproduce the average FIR/sub-mm SEDs for such galaxies remarkably well over a broad range of redshifts.  Main sequence galaxies make the dominant contribution to the star formation rate density at all epochs, and so this result adds confidence to the predictions of the model and the computation of the FIR SEDs of its galaxies.     
     
\section*{Acknowledgements}
The authors would like to thank Daniel Scharer for discussions that led to some of the developments in this paper, and Peter Mitchell, for providing us with the SED fitting code used in Appendix A.  We also acknowledge insightful comments from the referee, which allowed us to clarify many points made in the manuscript.  WIC and CGL acknowledge financial support and fruitful discussions from the Munich Institute for Astro- and Particle Physics (MIAPP) DFG cluster of excellence `Origin and Structure of the Universe' workshop entitled `The Star Formation History of the Universe', held in Munich in August 2015.  This work was supported by the Science and Technology Facilities Council [ST/K501979/1, ST/L00075X/1].  CMB acknowledges the receipt of a Leverhulme Trust Research Fellowship.  CL is funded by a Discovery Early Career Researcher Award (DE150100618). Parts of this research were conducted by the Australian Research Council Centre of Excellence for All-sky Astrophysics (CAASTRO), through project number CE110001020. This work used the DiRAC Data Centric system at Durham University, operated by the Institute for Computational Cosmology on behalf of the STFC DiRAC HPC Facility (www.dirac.ac.uk). This equipment was funded by BIS National E-infrastructure capital grant ST/K00042X/1, STFC capital grant ST/H008519/1, and STFC DiRAC Operations grant ST/K003267/1 and Durham University. DiRAC is part of the National E-Infrastructure. 
\bibliographystyle{mnras}
\bibliography{ref.bib}
\appendix
\section{The effect of a non-universal IMF on stellar masses inferred from SED fitting}
\label{app:IMF}
\begin{figure*}
\includegraphics[width = \linewidth]{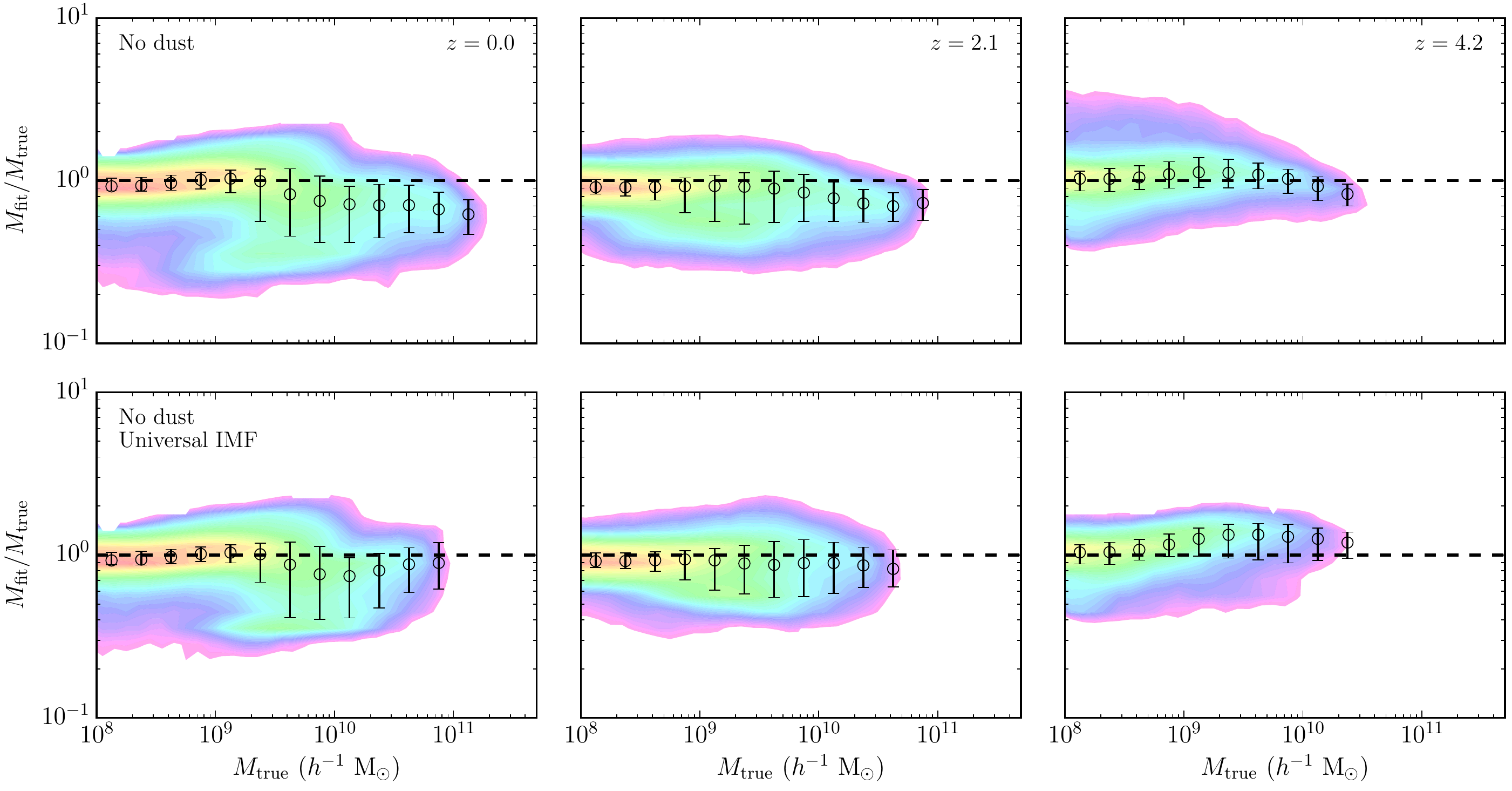}
\caption{The ratios of stellar masses inferred from broadband photometry using the SED fitting code presented in Mitchell et al. (\citeyear{Mitchell13}), ignoring attenuation by interstellar dust, to the true stellar masses predicted by \galform.  The colour scale indicates the  logarithmic density of points from red (high density) to purple (low density).  \emph{Top row:} Fiducial model with a top heavy IMF for burst mode star formation.  \emph{Bottom row:} Model with a universal IMF.  Open symbols and errorbars show the median and 16-84 percentiles of the distribution of inferred to true mass ratio at a given true stellar mass.  For reference, the horizontal black dashed line in each panel indicates unity.}
\label{fig:sed_fit_plot}
\end{figure*}

The galaxy formation model that we have used in this study (L16) incorporates two IMFs, a solar neighbourhood \cite{Kennicutt83} IMF for quiescent mode star formation, which occurs in the galactic disc, and a top-heavy IMF for burst mode star formation, which is triggered by some dynamical event and occurs in the galactic bulge. The top-heavy IMF is described by a slope of $x=1$ in $\mathrm{d}N(m)/\mathrm{d}\ln m\propto m^{-x}$ [for reference a \cite{Salpeter55} IMF has a slope of $x=1.35$].  Therefore galaxies in the model will contain stellar populations that formed with different IMFs.

Typically, stellar masses are inferred from observations by fitting model SEDs to observed broadband photometry making a number of assumptions \citep[for a discussion see e.g.][]{Pforr12,Mitchell13}, one of which is that the IMF is universal and has a form similar to that observed  for the solar neighbourhood.  Here we investigate what corrections, if any, it may be necessary to apply to the stellar masses predicted by the model to account for this assumption when comparing to observational data. 

To do this we use the SED fitting code presented in \cite{Mitchell13}.  We utilise simulated photometry from the same broadband filters as used in the \cite{Ilbert10} study that derived the stellar masses for the \cite{Bethermin15} sample we are comparing our model predictions to in this work.  These comprise $15$ bands, the \emph{GALEX} far- and near- UV, Subaru/SuprimeCam \emph{BVgriz}, CFHT/WIRCAM \emph{JHK} and the 4 \emph{Spitzer}/IRAC bands.  We also assume the same star formation history grid as used in \cite{Ilbert10}, that is a grid of exponentially decaying star formation histories, $\exp(-t_{\rm age}/\tau)$, where $\tau=0.1$, $1$, $2$, $3$, $5$, $10$, $15$, $30$~Gyr, and $t_{\rm age}$, the time since the star formation began, is constrained to be less than the age of the Universe.  We use the stellar population models of \cite{Maraston05}, which are calculated for a grid of 4 metallicities ($Z=0.02$, $0.5$, $1$, $2$~$Z_{\sun}$) and 67 ages ranging from $10^3$~yr to $15$~Gyr.  In the model SED fitting we always assume a universal \cite{Kennicutt83} IMF, and for simplicity we ignore the effects of dust attenuation.

We use SED fitting to derive (likelihood weighted) inferred masses for a sample of our model galaxies and show the ratio of inferred stellar mass to true stellar mass as a function of true stellar mass at a range of redshifts in Fig.~\ref{fig:sed_fit_plot}.  We do this for both the fiducial model (top row) and for a model which assumes a universal Kennicutt (\citeyear{Kennicutt83}) IMF\footnote{We do not consider this an acceptable model of galaxy formation as it fails to reproduce the observed number counts of galaxies at $850$~$\muup$m by more than an order of magnitude.} (bottom row).  

Even in the highly simplified case in which the effects of dust are ignored, there are no errors associated with the input photometry, and the same stellar population models and IMF are used in both the model and SED fitting (bottom row), the ratio of inferred to true stellar masses has a median value that can deviate from unity and shows $16-84$ percentile scatter of up to a factor of $\sim3$.  We note that the differences between the top and bottom rows in Fig.~\ref{fig:sed_fit_plot} that are caused by having a top-heavy IMF for burst mode star formation in the fiducial model are typically smaller than the amount of scatter seen in both rows and between different redshifts for the same model.  We conclude that any corrections due to having a non-universal IMF are small compared to the uncertainties associated with the SED fitting technique itself, and so would not have a significant effect on the results presented in this paper.  Therefore, we make no explicit correction for this in this study.  We caution however, that this may not be the case for some populations of galaxies, depending on the selection criteria, e.g. sub-mm galaxies selected by their $850$~$\muup$m flux.

\section{Comparing the simple dust model with \grasil}
\label{app:grasil_comparison}
\begin{figure*}
\includegraphics[width=\linewidth]{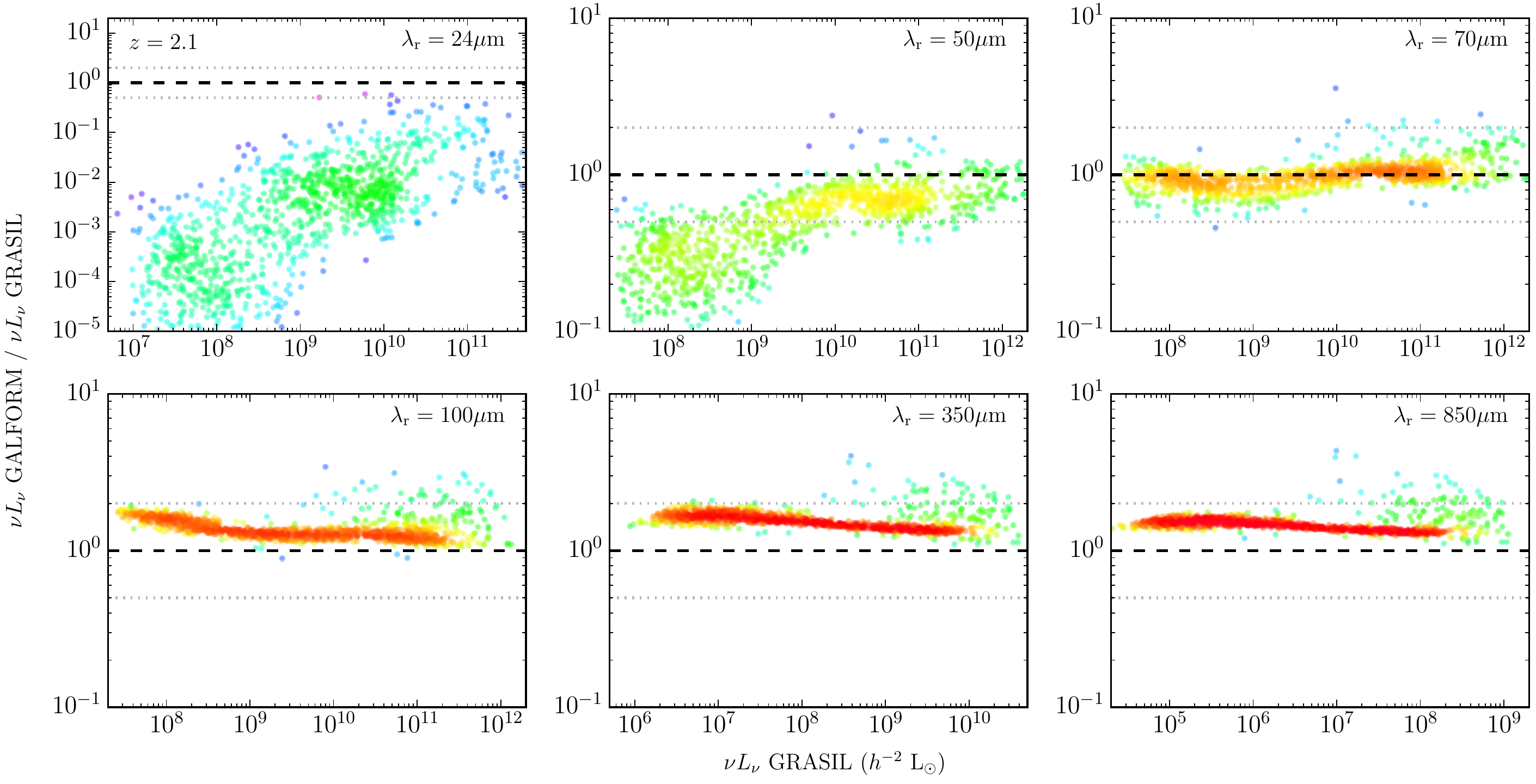}
\caption{Comparison of the predictions of the simple dust model with those from \grasil (Silva et al. \citeyear{Silva98}), at a range of rest-frame wavelengths $\lambda_{r}=24$, $50$, $70$, $100$, $350$ and $850$~$\muup$m, at redshift $z=2.1$, for a random sample of $\sim1000$ star-forming galaxies.  The black dashed line in each panel indicates unity and the grey dotted lines indicate a factor of $\pm0.3$~dex from unity.  Note that for the $\lambda_{\rm r}=24$~$\muup$m panel the ordinate axis covers a much larger dynamic range than for the other wavelengths considered.  The colourscale indicates the logarithmic density of points from red (high density) to purple (low density).}
\label{fig:grasil_scatter}
\end{figure*}

In this section we perform a brief comparison of the predictions made using our simple model for dust absorption and emission with those of the spectrophotometric radiative transfer code, \grasil \citep{Silva98}.  \grasil assumes a similar geometry to our simple dust model but treats some of the physics involved in more detail: (i) dust temperature calculated self-consistently at each location across the galaxy and also according to the size and composition of the dust grains (\grasil assumes a distribution of grain sizes and two compositions: graphite and astronomical silicate); (ii) temperature fluctuations for small grains due to finite heat capacities and (iii) the inclusion of PAH molecules.  Whilst the \grasil calculation is more physically sophisticated it is too computationally expensive to run for each galaxy in the samples we are considering in this work.  For this reason, we here restrict ourselves to a random sample of $\sim1000$ star-forming galaxies (as defined by sSFR$^{\prime}>$sSFR$^{\prime}_{\rm split}$) at $z\sim2$.  

A crude comparison of the computational cost of \grasil\ and \galform\ was made using the galaxy sample generated for this section. \grasil\ took, on average, $\sim160$~CPU~seconds per galaxy, while \galform\ (including the galaxy formation calculation) took $\sim6.2\times10^{-3}$~CPU~seconds of which approximately 10 per cent was spent computing SEDs.

Thus we infer that the \grasil\ calculation is $\sim2.6\times10^5$ longer than the simple dust model used here.
       
Where \grasil\ has analogous parameters to the simple dust model (e.g. $f_{\rm cloud}$, $t_{\rm esc}$, $\beta_{\rm b}$) used here for the purposes of this comparison we use the same values for each model.  Additional \grasil\ parameters are set to the values used by Baugh et al. (\citeyear{Baugh05}, see also Lacey et al. \citeyear{Lacey08}, Swinbank et al. \citeyear{Swinbank08} and Lacey et al. \citeyear{Lacey11}) without further recalibration. 

In Fig.~\ref{fig:grasil_scatter} we show the comparison of the luminosities predicted by the \galform\ dust model to those calculated using \grasil\ at rest-frame wavelengths of $24$, $50$, $70$, $100$, $350$ and $850$~$\muup$m, as indicated in the panels, at $z=2.1$.  We see that for rest-frame wavelengths of $\lambda_{\rm r}\gtrsim70$~$\muup$m the simple model can reproduce the results of \grasil\ to better than a factor of two (indicated in each panel by the grey dotted lines), with a relatively small amount of scatter.  However, at shorter rest-frame wavelengths the assumptions in the simple model break down, as can be seen by the increased scatter and larger deviation from unity.  We therefore restrict our comparisons of dust emission SEDs to rest-frame wavelengths longer than $70$~$\muup$m.

\bsp	
\label{lastpage}
\end{document}